\documentstyle[12pt]{article}

\makeatletter
\def\vereq#1#2{\lower3pt\vbox{\baselineskip1.5pt \lineskip1.5pt
\ialign{$\m@th#1\hfill##\hfil$\crcr#2\crcr\sim\crcr}}}
\makeatother

\begin{document}

\begin{titlepage}
\begin{center}
\today     \hfill    LBNL-40460\\
~{} \hfill UCB-PTH-97/38\\
~{} \hfill hep-th/9707133\\

\vskip .1in

{\large \bf Holomorphy, Rescaling Anomalies and Exact $\beta$
  Functions \\in Supersymmetric Gauge Theories}\footnote{This work
  was supported in part by the Director, Office of Energy Research,
  Office of High Energy and Nuclear Physics, Division of High Energy
  Physics of the U.S. Department of Energy under Contract
  DE-AC03-76SF00098 and in part by the National Science Foundation
  under grant PHY-90-21139.  NAH was also supported by NSERC, and HM
  by Alfred P. Sloan Foundation.}

\vskip 0.3in

Nima Arkani-Hamed and Hitoshi Murayama

\vskip 0.05in

{\em Theoretical Physics Group\\
     Ernest Orlando Lawrence Berkeley National Laboratory\\
     University of California, Berkeley, California 94720}

\vskip 0.05in

and

\vskip 0.05in

{\em Department of Physics\\
     University of California, Berkeley, California 94720}

\end{center}

\vskip .1in

\begin{abstract}
  There have long been known ``exact" $\beta$ functions for the gauge
  coupling in $N=1$ supersymmetric gauge theories, the so-called
  Novikov-Shifman-Vainshtein-Zakharov (NSVZ) $\beta$ functions.
  Shifman and Vainshtein further related these $\beta$ functions to the
  exact 1-loop running of the gauge coupling in a ``Wilsonian" action.
  All these results, however, remain somewhat mysterious. We attempt
  to clarify these issues by presenting new perspectives on the NSVZ
  $\beta$ function. Our interpretation of the results is somewhat
  different than the one given by Shifman and Vainshtein, having
  nothing to do with the distinction between ``Wilsonian" and ``1PI"
  effective actions.  Throughout we work in the context of the
  Wilsonian Renormalization Group; namely, as the cutoff of the theory
  is changed from $M$ to $M'$, we seek to determine the appropriate
  changes in the bare couplings needed in order to keep the low energy
  physics fixed.  The entire analysis is therefore free of infrared
  subtleties.  When the bare Lagrangian given at the cutoff is
  manifestly holomorphic in the gauge coupling, we show that the
  required change in the holomorphic gauge coupling is exhausted at
  1-loop to all orders of perturbation theory, and even
  non-perturbatively in some cases.  On the other hand, when the bare
  Lagrangian at the cutoff has canonically normalized kinetic terms,
  we find that the required change in the gauge coupling is given by
  the NSVZ $\beta$ function. The higher order contributions in the
  NSVZ $\beta$ function are due to anomalous Jacobians under the rescaling
  of the fields done in passing from holomorphic to canonical normalization.
  We also give prescriptions for regularizing certain $N=1$ theories
  with an ultraviolet cutoff $M$ preserving manifest holomorphy, starting from
  finite $N=4$ and $N=2$ theories. It is then at least in principle
  possible to check the validity of the exact $\beta$ function by
  higher order calculations in these theories.
\end{abstract}
\end{titlepage}

\newpage

\section{Introduction}
\setcounter{footnote}{0}
\setcounter{equation}{0}

In recent years, enormous progress has been made in understanding the
non-perturbative dynamics of supersymmetric gauge theories \cite{IS}.
Many theories have been solved ``exactly", setting the stage for
applications of strong supersymmetric gauge dynamics in building
realistic models of particle physics.

However, the connections between the exact results and those obtained
in perturbation theory are still not entirely clear.  One famous
example of a confusion in this regard is the anomaly puzzle. In supersymmetric
theories, the $U(1)_R$ current is in the same multiplet as the trace
of the energy-momentum tensor \cite{FZ}, and hence the chiral anomaly
and the trace anomaly are related \cite{CPS,PS}. The chiral anomaly is
exhausted at 1-loop \cite{AB}, however, implying that the trace
anomaly is exhausted also at 1-loop.  Since the trace anomaly
determines the gauge coupling $\beta$ function $\beta(g)$, this seems
to imply that this $\beta(g)$ should be also exhausted at 1-loop.
However, explicit perturbative calculations find higher order
corrections to $\beta(g)$ \cite{N=1}.

Shifman and Vainshtein \cite{SV} presented a solution to this puzzle,
by distinguishing between the ``Wilsonian" gauge coupling constant
$g_W$ and the ``1PI" or ``physical" coupling $g$.  In their
interpretation, $g_W$ appears in the Wilsonian effective action and
only runs at 1-loop, whereas $g$ appears in the 1PI effective action
and receives higher order corrections. Moreover, they presented a
remarkable formula relating the two types of gauge coupling from which
they obtained, to all orders in perturbation theory, the exact
$\beta(g)$ for $N=1$ supersymmetric gauge theories with matter fields
$\phi_i$. The same $\beta$ function was first derived via different
arguments by Novikov, Shifman, Vainshtein and Zakharov (NSVZ)
\cite{NSVZ}:
\begin{equation}
  \beta(g) = -\frac{1}{16 \pi^2} \frac{3 t_2(A) - \sum_i t_2(i)(1 -
    \gamma_i)}{1 - t_2(A)g^2/8 \pi^2} .
  \label{NSVZfull}
\end{equation}
Here $t_2(A),t_2(i)$ are the Dynkin indices for the adjoint and
$\phi_i$ representations (e.g. $t_2({\bf N})=1/2$ and $t_2(A)=N$ in
$SU(N)$), and $\gamma_i$ is the anomalous dimension of $\phi_i$.
Explicit perturbative calculations verify the NSVZ $\beta$ function up
to two-loop order.  This is clearly a significant achievement.  Beyond
two loops, however, the $\beta$ function coefficients are scheme
dependent, and it is not clear in what scheme the NSVZ $\beta$
function is supposed to be exact.\footnote{One can relate
  $\overline{\rm DR}$ scheme and the NSVZ scheme order by order in
  perturbation theory \cite{JJN}.}  This is one aspect of a general
confusion (which at least we have) surrounding the arguments leading
to the NSVZ $\beta$ function.

The purpose of this paper is to attempt to eliminate these confusions
by giving independent derivations of the NSVZ $\beta$ function.  Our
interpretation of the results, however, is somewhat different. We do
not use the 1PI effective action to define the gauge coupling
constant.  Instead, we work throughout in the context of the Wilsonian
Renormalization Group (WRG), which we briefly review
here.\footnote{For a general discussion of Wilsonian renormalization
  program in continuum field theories, see \cite{Wilson}.  See also
  \cite{AM} for a more complete discussion of the WRG invariance of
  exact results in SUSY gauge theories.} Any field theory is defined
with some cutoff $M$, and bare couplings $\lambda_0^i$. If we wish to
change the cutoff from $M$ to $M'$ while keeping the low energy
physics fixed (this step is often referred to as ``integrating out
modes between $M$ and $M'$''), we need to change the bare couplings
$\lambda^0_i \rightarrow \lambda^{0 \prime}_i$. The way in which the
$\lambda^0_i$ must change with the cutoff $M$ keeping the low energy
physics fixed is encoded in a Wilsonian Renormalization Group Equation
(WRGE) for the $\lambda^0_i$, $(Md/dM) \lambda^0_i =
\beta_i(\lambda^0)$.  All of the usual results of
renormalization-group analysis can be derived along these lines (see
\cite{AM} for some examples).  The virtue of this approach is the
freedom from infrared subtleties.  All the modes beneath $M'$ have yet
to be integrated over, so none of the calculations involve infrared
divergences.  Since the infrared effects are sensitive to the detailed
dynamics of different models, it is difficult to make exact and
non-trivial statements on the evolution of the coupling constant if the
calculation involves infrared effects.  By separating the infrared
physics from the discussion, we will be able to make concrete
statements on the ultraviolet structure of supersymmetric gauge
theories with confidence.  Having understood the ultraviolet
properties, the interesting physics lies in infrared non-perturbative
dynamics, which as we know can vary drastically depending on the
particular supersymmetric gauge theory under consideration.

In supersymmetric theories, we have two natural choices for the form
of the Lagrangian defined with cutoff $M$. The first is manifestly
holomorphic in the combination $1/g^2_h = 1/g^2 + i \theta/8 \pi^2$.
The second uses canonically normalized kinetic terms for all fields;
in this case the gauge coupling is called the canonical gauge coupling
$g_c$. We will show that, in changing the cutoff from $M$ to $M'$, the
change in $1/g^2_h$ needed to keep the low energy physics fixed is
exhausted at 1-loop, but that $g_c$ must be changed according to the
NSVZ $\beta$ function.  Furthermore, some special theories can be
explicitly regulated in a way that preserves holomorphy. The
validity of the exact $\beta$ function can then at least
in principle be checked by perturbative calculations in these theories.

The outline for the paper is as follows.  In Sec. 2, we consider pure
SUSY Yang-Mills theories, and show that the running of the holomorphic
gauge coupling $1/g^2_h$ is exhausted at 1-loop.  However, in the
rescaling of the vector multiplet needed to go to canonical
normalization, we encounter an anomalous Jacobian.  Correctly
accounting for this anomalous Jacobian gives the relation between
$g_c$ and $g_h$ given by Shifman and Vainshtein, and hence the exact NSVZ
$\beta$-function.   In Sec. 3, we address
the anomaly puzzle in our framework.  The resolution is very simple.
The anomaly under dilations (trace anomaly) is in the same multiplet
as the $U(1)_R$ anomaly and is one-loop exact.  However, because of
the anomaly, the vector multiplet does not have canonical kinetic
terms after the dilation.  If we wish to work with canonical kinetic
terms for the vector multiplet, a further change in normalization
(rescaling) of the vector multiplet must be done, which is itself
anomalous. Therefore, the anomaly from this ``modified" dilation
(naive dilation + rescaling of vector multiplet) is {\it not} in the
same multiplet as the $U(1)_R$ anomaly, and receives contributions
beyond 1-loop according to the NSVZ $\beta$ function.  In Sec. 4, we
extend the discussion on $\beta$ functions to the case with matter
fields. In Sec. 5 we consider $N=2$ theories, and use our results to
explain the finiteness of these theories beyond 1-loop.  In Sec.  6,
we give explicit prescriptions for regularizing 
some $N=1$ theories with a cutoff $M$, starting from finite $N=4$ and
$N=2$ theories. We explicitly define the couplings $g_h(M),g_c(M)$,
and show that the Shifman-Vainshtein relation holds between them. We
draw our conclusions in Sec. 7, while two appendices contain
discussions and explicit computations of all the required anomalous
Jacobians.

\section{Pure $N=1$ SUSY Yang--Mills}
\setcounter{footnote}{0}
\setcounter{equation}{0}

At a certain cutoff scale $M$, the Lagrangian for pure $N=1$
supersymmetric Yang-Mills (SUSY YM) can be given in two different
ways.  With the vector multiplet $V_h = V_h^a T^a$, we can write it in
a way that is manifestly holomorphic in the gauge coupling:
\begin{equation}
  {\cal L}^M_h(V_h) = \frac{1}{16}\int d^2 \theta \frac{1}{g_h^2} W^a(V_h)
  W^a(V_h) + \mbox{h.c.}
  \label{Lh}
\end{equation}
where $W^a_\alpha(V) T^a= \frac{-1}{4} \bar{D}^2 e^{-2V} D_\alpha e^{2
  V}$,\footnote{Our normalization of the vector multiplet differs from
  that of Wess and Bagger \cite{WB} by a factor of two, and we need to
  rescale it as $V_h = g_c V_c$ to go to canonical normalization.}
and
\begin{equation}
  \frac{1}{g_h^2} = \frac{1}{g^2} + i \frac{\theta}{8 \pi^2}.
  \label{gh}
\end{equation}
On the other hand, we can work with canonical normalization for the
gauge kinetic terms. In this case, the Lagrangian is written as
\begin{equation}
  {\cal L}^M_c(V_c) = \frac{1}{16}\int d^2 \theta \left(\frac{1}{g_c^2}
  + i \frac{\theta}{8 \pi^2}\right) 
  W^a(g_c V_c) W^a(g_c V_c) + \mbox{h.c.}
  \label{Lc}
\end{equation}
Note that since $g_c V_c$ is a real superfield, $g_c$ must be real,
and the Lagrangian is {\it not} holomorphic in the combination
$(1/g_c^2 + i \theta/8 \pi^2)$.

Suppose we now change the cutoff from $M$ to $M'$; how must the
couplings be changed to keep the low energy physics fixed? The answer
is particularly simple in the case of the holomorphic coupling. For
the holomorphic coupling at the cutoff $M'$, $1/g'^2_h$, let us write
\begin{equation}
  \frac{8 \pi^2}{g'^2_h} = \frac{8 \pi^2}{g^2_h} 
  + f\left(\frac{8 \pi^2}{g^2_h} , \ln \frac{M}{M'}\right).
\end{equation}
The function $f(8 \pi^2/g^2_h,t)$ must be holomorphic in $1/g^2_h$,
continuous in $t$, and must satisfy $f(8 \pi^2/g^2_h,0)=0$.  Since a
$2 \pi$ shift in $\theta$ has no effect, we must have $f(8 \pi^2/g^2_h
+ 2 \pi i,t) = f(8 \pi^2/g^2_h,t) + 2 \pi n(t) i$, where $n(t)$ is an
integer. However, since we know $n(0)=0$, by continuity in $t$, $n(t)
=0$.  Therefore $f(8 \pi^2/g^2_h + 2 \pi i,t) = f(8 \pi^2/g^2_h,t)$.
These observations can be cast in terms of the Wilsonian $\beta$
function for the holomorphic gauge coupling: we must have
\begin{equation}
  \frac{d}{dt} \left(\frac{8 \pi^2}{g^2_h}\right) = \beta \left(\frac{8
  \pi^2}{g^2_h}\right); \qquad
  \beta \left(\frac{8 \pi^2}{g^2_h} + 2 \pi i\right) = \beta
  \left(\frac{8 \pi^2}{g^2_h}\right) 
\end{equation}
Since the $\beta$ function is periodic, it can be Fourier expanded
\begin{equation}
  \beta \left(\frac{8 \pi^2}{g^2_h}\right) = 
  \sum_{n \geq 0} a_n e^{-n 8 \pi^2/g^2_h} 
\end{equation}
where the sum is restricted to $n \geq 0$ so that the theory makes
sense in weak coupling. The term with $n=0$ is a constant $a_0$ ($ = 
b_0$ is the one-loop $\beta$-function coefficient), and corresponds to
the 1-loop law for the running of $1/g^2_h$. The terms with $n \geq 1$
can never arise in perturbation theory.

In fact, for pure SUSY YM, a stronger argument shows that the terms
with $n \geq 1$ can not arise at all. Since the theory has an
anomalous $U(1)_R$ symmetry, if $1/g^2_h(t)$ is a solution to the
WRGE, $1/g^2_h + i \phi$ should also be a solution. This implies that
$\beta(8 \pi^2/g^2 + i (\theta + \phi)) = \beta(8 \pi^2/g^2 + i \theta)$, 
and hence that $\beta(8 \pi^2/g^2_h)$ is independent of 
$Im(8 \pi^2/g^2_h)$. 
However, any analytic function $f(z)$ which is independent of $Im(z)$ is
a constant.  Thus, the holomorphic $1/g^2_h$ runs {\it exactly} at
1-loop for pure SUSY YM, even including non-perturbative effects.

It is important to note that this result does not hold for other
definitions of $\beta$ functions. For instance, consider a pure
$SU(2)$ $N=2$ theory; this theory also has an anomalous $U(1)_R$
symmetry, so the argument given above implies that the running of
$1/g_h^2$ is exhausted at 1-loop. When the ($N=1$) adjoint chiral
superfield acquires a vev $\langle \phi \rangle = v \sigma_3$ breaking
$SU(2) \to U(1)$, Seiberg \cite{Seiberg} found that the effective
value of the holomorphic gauge coupling of the unbroken U(1) is given
by
\begin{equation}
  \frac{1}{g^2_{{\it eff}}(v)} = \frac{1}{g_h^2(M)} -\frac{b_0}{ 16 \pi^2}
  \mbox{ln}\frac{v^2}{M^2} +  c \left(\frac{M e^{8 \pi^2/b_0
  g_h^2(M)}}{v}\right)^4 + {\cal O}\left(e^{-16 \pi^2/g_h^2(M)}\right)
\end{equation}
where $b_0=-4$ is the coefficient of the 1-loop $\beta$ function, and
the constant $c$ as well as the higher order corrections have been
determined by Seiberg and Witten \cite{SW}.  If the $\beta$ function
is defined by $(v d/dv) \,g_{{\it eff}}(v) = \beta_{{\it eff}}(g_{{\it
    eff}})$, $\beta_{{\it eff}}$ contains both 1-loop and
non-perturbative corrections.  On the other hand, suppose we lower the
cutoff from $M$ to $M'$; how should $1/g^2_h(M)$ change to keep low
energy physics ($1/g_{{\it eff}}^2(v)$) fixed?  It is clear that the
required change in $1/g_h^2(M)$ is exhausted at 1-loop, i.e.
$1/g_h^2(M') = 1/g_h^2(M) - (b_0/8 \pi^2)$ln$M'/M$, and so the
Wilsonian $\beta$ function for the holomorphic gauge coupling is
indeed exhausted at 1-loop in this example.

We now wish to determine the Wilsonian $\beta$ function for the
canonical gauge coupling $g_c$. If we change the cutoff from $M$ to
$M'$, how must $1/g^2_c$ be changed to keep the low energy physics
fixed? At first sight, there seems to be no difficulty in going from
the holomorphic to canonical normalizations: simply making the change
of variable $V_h=g_c V_c$, the Lagrangian seems to have canonical
normalization 
for the vector multiplet with $g_c = g_h$. However, this is not
correct, as there is an anomalous Jacobian in passing from $V_h$ to
$V_c$; ${\cal D}(g_c V_c) \neq {\cal D} V_c$.\footnote{The measure
  ${\cal D}V$ is for the entire gauge sector of the theory, including
  the ghosts.}

In Appendix A, we explicitly compute this Jacobian, and at the end of
Sec.~5, we derive it indirectly based on the known finiteness of $N=2$
theories beyond 1-loop. The two methods yield the same result:
\begin{equation}
  {\cal D}(g_c V_c) = {\cal D}(V_c) \,\exp\left(\frac{1}{16}\int d^4 y 
    \int d^2 \theta \frac
    {2 t_2(A)}{8 \pi^2} \, \mbox{ln}g_c \, W^a(g_c V_c) W^a(g_c V_c)+
    \mbox{h.c.} + {\cal O}(1/M^4)\right)
  \label{JacV}
\end{equation}
where the $F$ terms given above are exact, and ${\cal O}(1/M^4)$
refers to higher dimension $D$ terms suppressed by powers of $1/M$
(the lowest dimension operator is of the form $\int d^4 \theta
WW\bar{W}\bar{W}/M^4$).

In a non-supersymmetric theory, it is not permissible to simply throw
away higher dimension operators suppressed by powers of the cutoff.
We can form relevant operators by closing the legs of higher dimension
operators, and power divergences in the loops can negate the cutoff
suppression of the higher dimension operators \cite{Wilson}.  As far
as physics at energy scales $E \ll M$ is concerned, however, the
Lagrangian with the higher dimension operators included yields the
same Green's functions\footnote{Up to corrections suppressed by powers
  in $(E/M)$.} as a different Lagrangian with all higher dimension
operators set to zero, but only after appropriately modifying the
coefficients of the relevant operators.  In our case, an important
simplification occurs: the usual supersymmetric non-renormalization
theorem \cite{non-renormalization} makes it impossible for the higher
dimension $D$ terms to ever produce an $F$ term such as $WW$, and
therefore no modification of the coefficient of $WW$ is needed upon
dropping the higher dimension $D$ terms.  Note also that since any
possible contribution of the higher dimension $D$ terms is coming from
ultraviolet divergences which need to negate the cutoff suppressions,
there is no worry about any subtle {\it infrared} singular $D$ terms
(such as $\int d^4 \theta W \frac{-D^2}{4\Box} W$) being generated.
This is welcome, since the non-renormalization theorem does not forbid
the generation of these operators \cite{NSVZ,SV,JW}, but they are
equivalent to $\int d^2 \theta WW$: $\int d^4 \theta W \frac{-D^2}{4
  \Box} W = \int d^2 \theta \frac{\bar{D}^2}{4} W \frac{D^2}{\Box} W =
\int d^2 \theta WW$ up to surface terms since $\bar{D}^2 D^2 W = 16
\Box W$.

With the Jacobian (\ref{JacV}), it is easy to derive the relationship
between the holomorphic and canonical gauge couplings (the
Shifman--Vainshtein formula \cite{SV}). At a fixed cutoff $M$, we
have\footnote {For compactness, we do not write the gauge-fixing terms
  in the path integrals which follow.}
\begin{eqnarray}
  Z&=&\int {\cal D}V_h \, \exp\left(-\frac{1}{16}\int d^4 y \int d^2 
    \theta \frac{1}{g^2_h} W^a(V_h) W^a(V_h) 
    + \mbox{h.c.} \right) \nonumber \\
  &=& \int {\cal D}(g_c V_c)\, \exp\left(-\frac{1}{16}\int d^4 y 
    \int d^2 \theta \frac{1}{g^2_h} 
    W^a(g_c V_c) W^a(g_c V_c) + \mbox{h.c.} \right) \nonumber \\
  &=& \int {\cal D}V_c\, \exp\left(-\frac{1}{16}\int d^4 y \int d^2 
    \theta \left(\frac{1}{g^2_h} - \frac{2 t_2(A)}{8 \pi^2}
    \mbox{ln}g_c \right)W^a(g_c V)W^a(g_c V) + \mbox{h.c.} \right).
\end{eqnarray}
In order to have canonical normalization for the vector multiplet, we
must have
\begin{equation}
  \frac{1}{g_c^2} = Re\left(\frac{1}{g_h^2}\right) 
  - \frac{2 t_2(A)}{8 \pi^2} \mbox{ln} g_c,
\label{svformula}
\end{equation}
which is the Shifman-Vainshtein formula. Since the difference between
$1/g_h^2$ and $1/g'^2_h$ is exhausted at 1-loop, we have
\begin{equation}
  \left(\frac{1}{g^{\prime 2}_c} + \frac{2 t_2(A)}{8 \pi^2} \mbox{ln}
    g'_c \right)  
  = \left(\frac{1}{g_c^2} + \frac{2 t_2(A)}{8 \pi^2} \mbox{ln} g_c
    \right) - \frac{3 t_2(A)}{8 \pi^2} \mbox{ln} \frac{M}{M'}. 
    \label{gcgpc}
\end{equation}  
The exact NSVZ $\beta$ function \cite{NSVZ} for pure SUSY YM then
follows trivially
\begin{equation}
  M \frac{d}{dM} g_c = \beta(g_c) = -\frac{3 t_2(A)}{16 \pi^2}
      \frac{g_c^3}{1 - \frac{t_2(A)}{8 \pi^2}g_c^2}.
      \label{NSVZbeta}
\end{equation}

It is noteworthy that the above derivation of the exact
$\beta$-function has no reference to 1PI effective actions or infrared
effects.  Indeed, the argument used here is exactly analogous to a
familiar argument on the chiral anomaly: the QCD Lagrangian with a
complex mass parameter
\begin{equation}
  {\cal L} = -\frac{1}{2} \mbox{Tr} F^{\mu\nu} F_{\mu\nu}
  + \bar{q} i {\not\!\!D} q + (m e^{i\phi} \bar{q}_{R} q_{L} +
  \mbox{h.c.})
  \label{eq:QCD1}
\end{equation}
can be brought to a Lagrangian with a real mass
\begin{equation}
  {\cal L} = -\frac{1}{2} \mbox{Tr} F^{\mu\nu} F_{\mu\nu}
  + \bar{q} i {\not\!\!D} q + (m \bar{q}_{R} q_{L} + \mbox{h.c.})
  + \frac{i\phi}{16 \pi^{2}} \mbox{Tr} F^{\mu\nu} \tilde{F}_{\mu\nu} .
  \label{eq:QCD2}
\end{equation}
In this case, the mass parameter is supposed to be the bare mass with
a fixed ultraviolet cutoff.  These Lagrangians describe exactly the
same low-energy physics.  The situation with the 1PI effective action
is more complicated, requiring a detailed discussion on how the
low-energy effective $\theta$ parameter is related to the bare
$\theta$-parameter \cite{theta-eff}.  However, one does not need to
worry about subtleties concerning infrared effects as long as one is
dealing with the change of bare parameters needed to keep the physics
fixed as the ultraviolet cutoff is varied, because the modes beneath
the cutoff are still to be integrated over.  Even though the bare
parameters are not as directly related to physical observables as
those in 1PI effective actions, making exact statements about the
physical equivalence of bare Lagrangians such as (\ref{eq:QCD1}) and
(\ref{eq:QCD2}) is crucial in many applications: {\it e.g.}\/
the determination of the effective Lagrangians for SUSY YM and SUSY QCD
theories in \cite{TVY}. 

The fact that we do not refer to the 1PI effective action is
desirable.  In non-abelian gauge theories, the infrared effects are so
severe that 1PI effective action cannot be defined without a clear
prescription for an infrared cutoff.  In fact, it is not clear what
the correct definition of the running gauge coupling constant is in
1PI effective actions.  One obvious choice is to use dimensional
regularization (or dimensional reduction), which regularizes both the
ultraviolet and infrared, possibly with minimal subtraction.
Dimensional regularization, however, is not desirable for our purposes
precisely because it regularizes both the ultraviolet and infrared
divergences; it is impossible to only move the ultraviolet cutoff
while leaving the infrared cutoff fixed, and hence it is hard to
disentangle different effects.  Actually, there is no rescaling
anomaly when dimensional regularization is used (see appendix A.1).
The two-loop contribution to the $\beta$-function of gauge coupling
constant, which we describe as a consequence of the rescaling anomaly,
appears from infrared uncertain terms $\sim 0/0$ \cite{0/0} in
perturbative calculations, when dimensional regularization is used.
This let the authors of Refs.~\cite{NSVZ2,SV} claim that the
$\beta$-function beyond one-loop arises from the infrared in SUSY YM and
supersymmetric QED.  However, it is not clear from this argument that
the two-loop contribution is due to infrared effects, since
dimensional regularization mixes up infrared and ultraviolet effects.
In fact, in the method of differential renormalization
\cite{differential}, it is clear that conventional $\beta$-functions
come only from short distance divergences, and the method reproduces
standard results for the 2-loop $\beta$-functions of Wess--Zumino
model \cite{WZ} and supersymmetric QED \cite{SQED}.  

It is nevertheless interesting to ask how the bare coupling
constant is related to gauge coupling constant in 1PI effective
actions.  Recall first that the coupling constants in 1PI effective
actions are highly scheme dependent (even gauge-dependent).  In order
to relate the Wilsonian coupling (holomorphic or canonical) to the 1PI
coupling, the renormalization scheme must be completely specified.  We
cannot make a general statement relating Wilsonian and 1PI couplings.
However, we expect that the canonical gauge coupling is closely
related to the 1PI coupling.  For instance, one can define the 1PI
coupling $g_{\rm 1PI}(q^2)$ by the gauge field two-point function at a
fixed Euclidean momentum transfer $q^2$ within the background field
formalism.  By changing the cutoff down to a scale very close to
$q^2$, one can minimize the difference between the bare Lagrangian and
the classical 1PI effective action for external fields of momentum
$O(q^2)$, since the path integral over quantum fields generates little
correction.  Therefore the 1PI coupling should be very close to the
canonical coupling, $g_{\rm 1PI} (q^2) \approx g_c (q^2)$.  If one starts
with the bare action with holomorphic normalization, the path integral
over quantum fields is not trivial even when $M^2 \approx q^2$,
because they do not have canonical normalization; the path integral
yields the difference between the holomorphic and canonical coupling,
and hence one obtains the same result as in the canonical case.  Even
though the argument in this paragraph is certainly not rigorous, it
does suggest the 1PI coupling is related to $g_c$ rather than $g_h$.
Calling $g_c$ the ``1PI coupling'' may not be incorrect; the statement
about an exact $\beta$-function is, however, somewhat empty unless a
renormalization scheme is specified.  There may also be
non-perturbative corrections from the path integral which cannot be
seen from this type of perturbative calculation.

\section{Anomaly Multiplet}
\setcounter{footnote}{0}
\setcounter{equation}{0}

One of the confusing points relating to $\beta$-functions in $N=1$
pure SUSY YM is the so-called anomaly multiplet puzzle.  At the
classical level, the $U(1)_{R}$ current belongs to the same multiplet
as the energy-momentum tensor (the supercurrent multiplet) \cite{FZ}.
Their anomalous divergences also form the chiral ``anomaly multiplet"
\cite{CPS,PS}, whose $F$ component is nothing but
\begin{equation}
  F = \frac{2}{3} \theta_\mu^\mu + i \partial_\mu j^\mu_R;
\end{equation}
holomorphy relates the $U(1)_R$ and trace anomalies.

The supersymmetric extension of Adler--Bardeen theorem states that the
anomaly of $U(1)_{R}$ is exhausted at one-loop \cite{AB}.  On the
other hand, the trace of the energy momentum-tensor is proportional to
the $\beta$-function of the gauge coupling constant, and hence
receives all order contributions.  This has been referred to as the
anomaly puzzle in supersymmetric gauge theories.  Grisaru, Milewski
and Zanon \cite{Ohat} studied this question in detail and found that
there are two different definitions of the supercurrent.  One
definition has anomaly exhausted at one-loop and belongs to the same
multiplet as the Adler--Bardeen $U(1)_{R}$ anomaly; the other has
anomaly from all orders in perturbation theory and is proportional to
the $\beta$-function.  The two operators were defined by
regularization via dimensional reduction, and differ in the $\epsilon$
dimensions.  Even though this could well be the resolution of the
puzzle, the discussion is highly technical, and the physical meaning
of the two operators is not clear. Shifman and Vainshtein \cite{SV}
also presented a solution to the anomaly puzzle.  In their
interpretation, the operator equation for the anomalies are indeed
exhausted at 1-loop. The all-order contribution to the trace anomaly
comes from infrared singularities which arise upon taking the matrix
element of the operator relations.  However, having understood the
NSVZ $\beta$ function in a purely Wilsonian framework with no
reference to infrared physics in the previous section, we would now like
to address the anomaly puzzle in our framework.

In our language, the resolution to the puzzle is very simple.  The
anomaly under the $U(1)_{R}$ transformation and dilation are related
by holomorphy.  As long as one maintains the manifest holomorphy, they
have anomalies in the same multiplet, and are exhausted at 1-loop.  
On the other hand, if the
vector multiplet has canonical kinetic term, it will not stay
canonical after the dilation has been performed, and an additional
rescaling is needed in order to go back to canonical normalization.
Therefore, this modified dilation (which keeps canonical normalization
for the vector multiplet) is no longer related to the $U(1)_R$
transformation, and its anomaly receives contributions from all orders
in perturbation theory according to the NSVZ $\beta$ function.  The
two different definitions of the trace of energy momentum tensor are
consequences of the two different dilation transformations: the naive
one appropriate in holomorphic normalization and the modified one
which is designed to preserve canonical normalization.  We do not work
out the explicit forms of the energy momentum tensor here; instead we
explain what dilation transformations are appropriate for the two different
normalizations of the vector multiplet and explain how their anomalies
naturally differ.

The anomalous Jacobian under the dilation is worked out in Appendix B.
It is given by
\begin{equation}
  {\cal D}(V_h(e^{-t} x, e^{-t/2}\theta, e^{-t/2}\bar{\theta}))
  = {\cal D}V_h(x, \theta, \bar{\theta}) \,
      \exp \left( \frac{1}{16} 
        t \int d^{2}\theta \frac{-3 t_{2}(A)}{8\pi^{2}} W W +
  \mbox{h.c.} + O(1/M^{4}) \right) ,
\label{JacVdil}
\end{equation}
and the $F$-terms are exact just as was the case with the rescaling
anomaly.  This Jacobian adds to the gauge kinetic term and changes the
gauge coupling constant.  It is nothing but the required change in the
holomorphic gauge coupling constant under the change of the cutoff,
\begin{equation}
        \frac{1}{g_{h}^{2}} \rightarrow \frac{1}{g_{h}^{2}} + 
        \frac{b_{0}}{8\pi^{2}} t,
\end{equation}
as $b_{0} = - 3 t_{2}(A)$ in $N=1$ pure Yang--Mills theory.  The
exactness of (the $F$-term in) the anomalous Jacobian is in one-to-one
correspondence to the one-loop nature of the running of the
holomorphic gauge coupling constant.  It is also clear that the
Jacobians under $U(1)_{R}$ and dilation are given in a holomorphic
way, both proportional to the $WW$ operator.  This is nothing but the
anomaly multiplet structure, namely that the divergence of the
$U(1)_{R}$ current and the trace of the energy momentum tensor are
both given by $\int d^{2} \theta WW$ operator.  An explicit
regularization method which preserves the manifest holomorphy between
$U(1)_R$ and trace anomalies will be presented in Sec. 6.

The resolution to the anomaly puzzle is the normalization of the
vector multiplet.  Under the dilation, the gauge kinetic term receives
an additional contribution from the Jacobian.  When one employs
holomorphic normalization for the vector multiplet, this is the
correct dilation, and no further steps are necessary.  On the other
hand, starting with a canonically normalized vector multiplet, the
additional contribution to the quantum action from the anomalous
Jacobian (\ref{JacVdil}) takes the vector multiplet out of canonical
normalization.  Therefore, one needs to rescale the vector multiplet
to go back to canonical normalization, and this produces another
anomalous Jacobian.  We have the following sequence for the change in
the gauge kinetic term.  Starting with gauge
kinetic term in canonical normalization:
\begin{equation}
        \frac{1}{16} \int d^{2} \theta 
                \frac{1}{g_{c}^{2}} W^a(g_{c} V_{c}) W^a(g_{c} V_{c}),
\end{equation}
the dilation generates an additional contribution to the 
kinetic term, yielding
\begin{equation}
        \frac{1}{16} \int d^{2} \theta 
                \left(\frac{1}{g_{c}^{2}} + \frac{b_{0}}{8\pi^{2}} t\right)
                        W^a(g_{c} V_{c}) W^a(g_{c} V_{c}) .
\end{equation}
But now the vector multiplet is no longer canonically normalized.  A
modified dilation for the vector multiplet which would keep it in
canonical normalization is not only the transformation defined above
but further requires a rescaling of the vector multiplet. The change
of variable $g_{c} V_{c} = g'_{c} V'_{c}$ produces an additional
Jacobian as in the previous section, and the gauge kinetic term becomes
\begin{equation}
  \frac{1}{16} \int d^{2} \theta 
  \left(\frac{1}{g_{c}^{2}} + \frac{3 t_{2}(A)}{8\pi^{2}} t
    -2 \frac{t_{2} (A)}{8\pi^{2}} \ln \frac{g'_{c}}{g_{c}} \right)
  W^a(g'_{c} V'_{c}) W^a(g'_{c} V'_{c}). 
\end{equation}
Since this modified dilation
\begin{equation}
  V'_c(x, \theta, \bar{\theta}) = \frac{g_c}{g'_c} 
  V_c (e^{-t} x, e^{-t/2}\theta, e^{-t/2}\bar{\theta}) 
\end{equation}
includes the rescaling of the vector multiplet, it is not in the same
anomaly multiplet as the $U(1)_{R}$ transformation.  Now, $g'_{c}$
must be chosen so that the coefficient of the $WW$ operator becomes
$1/g'^{2}_{c}$, giving
\begin{equation}
  \frac{1}{g^{\prime 2}_c} = \frac{1}{g_c^2} 
  + \frac{3 t_2(A)}{8 \pi^2} t - \frac{2 t_2(A)}{8 \pi^2} 
  \mbox{ln} \frac{g_c'}{g_c}. 
\label{dilgc}
\end{equation}
Of course, in performing the dilation, the cutoff is changed from $M$
to $M'=e^t M$, so $1/g^{\prime 2}_c$ is the canonical bare coupling
needed at cutoff $M'$ in order to keep the physics fixed.
Eq.~(\ref{dilgc}) then gives the relationship between $g_c$ and $g'_c$
\begin{equation}
  \left(\frac{1}{g^{\prime 2}_c} + \frac{2 t_2(A)}{8 \pi^2} \mbox{ln}
    g'_c \right) 
  = \left(\frac{1}{g_c^2} + \frac{2 t_2(A)}{8 \pi^2}
    \mbox{ln} g_c \right) - \frac{3 t_2(A)}{8 \pi^2} \mbox{ln} \frac{M}{M'},
\end{equation}
which is just Eq.~(\ref{gcgpc}), and the NSVZ $\beta$ function follows
as in Eq.~(\ref{NSVZbeta}).

\section{Models with Matter Fields}
\setcounter{footnote}{0}
\setcounter{equation}{0}

In the case with matter multiplets, the holomorphic Lagrangian at
cutoff $M$ is
\begin{equation} 
  {\cal L}^M_h(V_h,\phi) = \frac{1}{16}\int d^2 \theta 
  \frac{1}{g_h^2} W^a(V_h)W^a(V_h) + \mbox{h.c.} 
  + \int d^4 \theta \sum_i \phi_i^{\dagger} e^{2 V^i_h} \phi_i,
\end{equation} 
where $i$ runs over the chiral multiplets and $V_h^i = V_h^a T^a_i$
($T^a_i$ are the generators in the $i$ representation).  There are
hidden parameters in the above Lagrangian, the coefficients $Z_i$ of
the kinetic terms for the chiral multiplets. However, we have made the
conventional choice and set $Z_i=1$ in the bare Lagrangians. Now, as
we change the cutoff from $M$ to $M'$, how must the couplings change
in order to keep the low energy physics fixed? Exactly the same
argument as in the Sec. 2 shows that, as long as the change in
$1/g^2_h$ is holomorphic, this change is exhausted at 1-loop.
However, the change can only be holomorphic if we allow the
coefficient of the matter kinetic terms (which are manifestly
non-holomorphic, being only a function of $Re(1/g^2_h)$) to change
from 1 to $Z_i(M,M')$, so that the Lagrangian at cutoff $M'$ is
\begin{eqnarray}
  {\cal L}^{M'}_h(V_h,\phi) &=& 
  \frac{1}{16} \int d^2 \theta \left(\frac{1}{g_h^2} + \frac{b_0}{8 \pi^2}
    \mbox{ln}\frac{M}{M'}\right)  
  W^a(V_h)W^a(V_h) + \mbox{h.c.} \nonumber \\ 
  &+&  \int d^4 \theta 
  \sum_i Z_i(M,M') \phi_i^{\dagger} e^{2 V^i_h} \phi_i 
\end{eqnarray}
where $b_0 = -3 t_2(A) + \sum_i t_2(i)$.  If we insist on working with
canonically normalized kinetic terms for the matter fields, we need to
make the change of variable $\phi'_i = Z_i(M,M')^{-1/2} \phi_i$. As
with the vector multiplet, however, the measure is not invariant under
this change, and there is an anomalous Jacobian \cite{KS}. In our
case, $Z_i(M,M')$ is real, but it is sensible to look at ${\cal
  D}(Z^{-1/2} \phi')$ for a general complex $Z$ since $\phi'$ is a
chiral superfield. Note that when $Z=e^{i \alpha}$ is a pure phase,
the change of variable is just a phase rotation of all the components
of $\phi$, and the Jacobian is just the one associated with the chiral
anomaly. This Jacobian is exactly known and cutoff independent
\begin{equation}
  {\cal D}(e^{-i \alpha/2} \phi') {\cal D}(e^{+i \alpha/2} \bar{\phi}')
  = {\cal D}\phi' {\cal D}\bar{\phi}'\, \exp\left(\frac{1}{16}
    \int d^4 y \int d^2 \theta \frac{t_2(\phi)}{8 \pi^2}
  \mbox{ln}(e^{i \alpha})  WW + \mbox{h.c.}\right).
\end{equation}
In the case where $Z$ is a general complex number, the cutoff
independent piece of this Jacobian has been calculated in a manifestly
supersymmetric way by Konishi and Shizuya \cite{KS}, and we present a
less technical derivation using components in the appendix. In general
the Jacobian has both $F$ and $D$ terms. The $F$ terms are known
exactly and are the same as in the above with ln$e^{i \alpha}$
replaced by ln $Z$. The $D$ terms (such as $Re($ln$Z)\bar{W} \bar{W} W
W$) are all suppressed by powers of the cutoff, and can be truly
neglected for the same reason as given in the Sec. 2: the
non-renormalization theorem makes it impossible for these operators to
contribute to $F$ terms.

Therefore, if we wish to work with canonically normalized matter
fields at all cutoffs, the Lagrangian at cutoff $M'$ must be
\begin{equation}
  {\cal L'}^{M'}_h(V_h,\phi) = \frac{1}{16} \int d^2 \theta \frac{1}{g'^2_h}
  W^a(V_h)W^a(V_h) + \mbox{h.c.} 
  + \int d^4 \theta \sum_i \phi^{\dagger} e^{2 V_h^i} \phi
\end{equation}
with
\begin{equation}
  \frac{1}{g'^2_h} = \frac{1}{g^2_h} + \frac{b_0}{8 \pi^2} \mbox{ln}
  \frac{M}{M'} - \sum_i \frac{t_2(i)}{8 \pi^2} \mbox{ln}Z_i(M,M').
\end{equation}
If we now wish to further work with canonical kinetic terms for the
vector multiplet, we need to rescale the vector field as in the
previous section, with the same result. The combination
\begin{equation}
  \frac{1}{g_c^2} + \frac{2 t_2(A)}{8 \pi^2} \mbox{ln} g_c 
  + \sum_i \frac{t_2(i)}{8 \pi^2} \mbox{ln}Z_i
\end{equation}
runs only at 1-loop, and the NSVZ $\beta$ function (\ref{NSVZfull})
follows trivially
\begin{equation}
  \mu \frac{ d g_c}{d \mu}  = -\frac{g_c^3}{8 \pi^2} 
  \frac {3 t_2(A) - \sum_i t_2(i) (1 - \gamma_i)}
  {1 - t_2(A) g_c^2/8 \pi^2}
\end{equation}
where $\gamma_i = (\mu d/d\mu)$ln$Z_i(M,\mu)$.

\section{$N=2$ Theories}
\setcounter{footnote}{0}
\setcounter{equation}{0}

We now turn to the analysis of $N=2$ theories, which are known to be
finite above 1-loop \cite{N=2}. Here we will explain this result by
using the anomalous Jacobians we have derived. We can also use the
known finiteness of these theories above 1-loop, proved
perturbatively, to give an alternate derivation of the Jacobian for
the rescaling of the vector multiplet, which we used to derive the
Shifman-Vainshtein formula (\ref{svformula}).

Using $N=1$ language, the holomorphic Lagrangian for pure $N=2$
supersymmetric Yang-Mills theories is
\begin{equation}
  {\cal L}(V_h,\phi_h) = \frac{1}{16}\int d^2 \theta 
  \frac{1}{g^2_h} W^a(V_h) W^a(V_h) + 
  \mbox{h.c.} 
  + \int d^4 \theta Re
  \left(\frac{2}{g^2_h}\right) \mbox{Tr} \phi_h^{\dagger}e^{-2
    V_h}\phi_he^{2 V_h} 
\end{equation}   
where $\phi_h$ is a chiral field in the adjoint representation of the
gauge group. As discussed in the previous sections, the holomorphic
gauge coupling only changes at 1-loop when we change the cutoff from
$M$ to $M'$.  The coefficients of the kinetic terms of the $N=1$ 
vector multiplet and the adjoint field are both changed according to 
the holomorphic gauge coupling as required by $N=2$ invariance.
If we wish to work with canonically normalized fields, we
must make the change of variables $\phi_h = g_c \phi_c,V_h=g_c V_c$
(where the rescalings for $\phi,V$ must be the same by $N=2$
supersymmetry). We can compute the Jacobian for this variable change
from the Jacobians for matter and vector fields we found in the
previous two sections.  The final result is that the Jacobian for the
vector multiplet cancels the one from the adjoint chiral multiplet:
\begin{eqnarray}
  {\cal D}(g_c V_c) &=& {\cal D} V_c\, \exp\left(\frac{1}{16}\int d^4 y 
    \int d^2 \theta \frac
    {2 t_2(A)}{8 \pi^2} \mbox{ln}g_c W^a(g_c V_c) W^a(g_c V_c)+ \mbox{h.c.} + 
    {\cal O}(1/M^4)\right) , \nonumber \\
  {\cal D}(g_c \phi_c) &=& {\cal D} \phi_c \,\exp
  \left(\frac{1}{16}\int d^4 y 
    \int d^2 \theta - \frac
    {2 t_2(A)}{8 \pi^2} \mbox{ln}g_c W^a(g_c V_c) W^a(g_c V_c)+ \mbox{h.c.} + 
    {\cal O}(1/M^4)\right) . \nonumber \\
  &&
\end{eqnarray}
Therefore the canonical coupling coincides with the holomorphic one,
and so pure $N=2$ theories must be perturbatively finite above 1-loop.
When $N=2$ hypermultiplets are added to the theory, the $\beta$
function still vanishes above 1-loop, since the kinetic terms for the
hypermultiplets are not renormalized \cite{NoZ}, so there is no need
to rescale them to go back to canonical normalization.
 
As already mentioned, we can turn around the above arguments. Since
the finiteness of $N=2$ theories beyond 1-loop has been explicitly
established in perturbation theory, it must be that the canonical
coupling coincides with the holomorphic coupling for these theories,
which in turn means that ${\cal D}(g_c V_c) {\cal D}(g_c \phi_c)={\cal
  D}V_c{\cal D}\phi_c$. However by holomorphy, the Jacobian for ${\cal
  D}(g_c \phi_c)$ can be inferred from the chiral anomaly Jacobian
$D(e^{i\alpha} \phi_c)$ without computation. From this, we can deduce
the Jacobian for the vector multiplet, and therefore the
Shifman-Vainshtein formula, as we did in Sec. 3.

\section{Regularization}
\setcounter{footnote}{0}
\setcounter{equation}{0}

In all the above analysis, we have somewhat loosely been referring to
the theory defined with a cutoff $M$, without defining how the theory
is to be cut off.  This problem is also related to the question of the
scheme in which the Shifman-Vainshtein formula, and hence the NSVZ
$\beta$-function, holds. We address these questions in this section,
by explicitly regulating some $N=1$ theories using finite $N=4$ and
$N=2$ theories. We will then give an explicit definition of
$1/g^2_h(M)$ and $1/g_c^2(M)$, and will show that they are related by
the Shifman-Vainshtein formula.

The idea is very simple. Let us begin with theories with $N=4$
supersymmetry, which are known to be finite \cite{N=4}. In $N=1$
language, these theories contain 1 vector multiplet $V$ and 3 chiral
multiplets $\phi^i$ in the adjoint representation.  Now, suppose we
add a mass term $\int d^2\theta M$Tr$(\phi^i \phi^i) +$
h.c. to the adjoints. Since this is a soft breaking of $N=4$ SUSY, the
theory is still free of UV divergences \cite{softN4}. Beneath the
scale $M$, it looks like pure $N=1$ SYM.  Thus, we have an explicit
regularization for pure $N=1$ SUSY YM with a cutoff $M$, and the
cutoff moreover preserves manifest holomorphy.

To be specific, we define the holomorphic pure $N=1$ SUSY YM theory, 
regulated with cutoff $M_h$ and with gauge coupling $1/g_h^2(M)$, by the 
Lagrangian
\begin{eqnarray}
  {\cal L}^{M_h}_h(V_h,\phi_{h}^i)&=&\frac{1}{16} \int d^2 \theta
  \frac{1}{g^2_h(M_h)}  W^a(V_h)W^a(V_h) + 
  \mbox{h.c.} \nonumber \\
  &+& \int d^4 \theta
  Re\left(\frac{2}{g_h^2(M_h)}\right) 
  \mbox{Tr}(\phi_{h}^{i\dagger} e^{-2 V_h} \phi_{h}^i e^{2 V_h}) 
  \nonumber \\
  &+& \int d^2 \theta Re\left(\frac{1}{g_h^2(M_h)}\right)
  \sqrt{2} \mbox{Tr}
  (\phi_{h}^i [\phi_{h}^j,\phi_{h}^k]) \frac{\epsilon_{ijk}}{3!} 
  + M_h \mbox{Tr} 
  (\phi_{h}^i \phi_{h}^i) + \mbox{h.c.} 
\end{eqnarray}
Since this theory is finite, its ultraviolet cutoff is taken to be
infinite.  The coupling $1/g_h^2 (M_h)$ is the holomorphic coupling of
the theory with the infinite ultraviolet cutoff and is finite.  On the
other hand, we are interested in this theory as the regularized $N=1$
SUSY YM with ultraviolet cutoff $M_h$.  The holomorphic coupling
$1/g_h^2 (M_h)$ can be specified independently of $M_h$; we specify
$M_h$ as the argument, however, because later in
Eq.~(\ref{eq:Mh-independence}) we will vary $M_h$ and $1/g_h^2 (M_h)$
at the same time keeping correlation functions of the $N=1$ SUSY YM
fixed.

We similarly define the theory with canonically normalized kinetic terms for
the vector multiplet, regulated with cutoff $M_c$ and with gauge coupling
$1/g_c^2$ by 
\begin{eqnarray}
  {\cal L}^{M_c}_c(V_c,\phi_{c}^i)&=& \frac{1}{16} \int d^2 \theta 
  \frac{1}{g^2_c(M_c)} W^a(g_c(M_c)V_c)
  W^a(g_c(M_c)V_c) + 
  \mbox{h.c.} \nonumber \\
  &+& \int d^4 \theta\,
  2 \mbox{Tr} (\phi_{c}^{i\dagger} e^{-2 g_c(M_c)V_c} \phi_{c}^i e^{2
    g_c(M_c)V_c})  
  \nonumber \\
  &+& \int d^2 \theta \, \sqrt{2} g_c (M_c)
  \mbox{Tr}(\phi_{c}^i[\phi_{c}^j,\phi_{c}^k]) 
  \frac{\epsilon_{ijk}}{3!} 
  + M_c \mbox{Tr}(\phi_{c}^i \phi_{c}^i) 
  + \mbox{h.c.}
\end{eqnarray}
where the relative normalizations of all the terms in the above are chosen to
ensure $N=4$ supersymmetry in the $M_c \rightarrow 0$ limit. 

The first thing we have to check is that, as $M_h$ is changed, the
holomorphic coupling in the $N=4$ Lagrangian must be changed according
to the one-loop law to keep the low-energy physics fixed. 
Requiring that correlation functions of low-energy fields 
do not vary as $M_h$ is changed, we have\footnote{Up to 
corrections suppressed by powers in $1/M_{h}$.}
\begin{equation}
  M_h \frac{d}{d M_h} \int {\cal D} V_h \prod_{i=3}^3 {\cal D}\phi^i_h
  e^{-S}
  {\cal O}_1 \cdots {\cal O}_n = 0,
\label{eq:Mh-independence}
\end{equation}
and its complex conjugate equation in terms of $\bar{M}_h$.  Here and
below, ${\cal O}_i$ are arbitrary operators of $V_h$, and $S = \int
d^4 x {\cal L}^{M_h}_h(V_h,\phi_{h}^i)$.  The change
required for $1/g_h^2(M)$ is therefore determined by the equation
\begin{equation}
  \int {\cal D} V_h \prod_{i=3}^3 {\cal D}\phi_h^i e^{-S}
  {\cal O}_1 \cdots {\cal O}_n
  \left\{ \left(M_h \frac{d}{d M_h} \frac{1}{g_h^2 (M_h)}\right) S_0
  + \int d^4 x d^2 \theta M_h \mbox{Tr} \phi^i_h \phi^i_h \right\} = 0,
\label{eq:change1}
\end{equation}
where $S_0$ is the action without the mass term and with the overall
$1/g_h^2(M_h)$ dropped.  Recall that the operators ${\cal O}_i$ of
physical interest do not involve the regulator fields $\phi^i$.
Therefore, we can replace the $M_h \mbox{Tr} \phi_h^i \phi_h^i$
operator in the above matrix element by operators of the low-energy
fields, {\it i.e.}\/ $W_\alpha$, in the sense that any correlation
function of the above operator $\int d^4 x d^2 \theta M_h \mbox{Tr}
\phi^i \phi^i$ with other operators of low-energy fields can be given,
through a systematic expansion in $1/M_h$, by operators which involve
low-energy fields only.\footnote{In other words, we take the
  expectation value of $M_h \mbox{Tr} \phi^i_h \phi^i_h$ within ${\cal
    D}\phi_h^i$ path integral in the background gauge field $V_h$,
  expanding in powers of $1/M_h$.  This is the valid procedure because
  none of the operators ${\cal O}_i$ depend on $\phi_h^i$ and are
  outside the ${\cal D}\phi_h^i$ path integral.}
Such an expansion can be done 
easily\footnote{This calculation is identical to the derivation of the
  chiral anomaly \cite{Jackiw} or the
  Konishi anomaly \cite{KS} with Pauli-Villars regularization, 
  the only difference being the opposite statistics of the 
  $\phi_i$ relative to the Pauli-Villars regulator fields.} and
yields
\begin{eqnarray}
  \lefteqn{
\langle {\cal O}_1 \cdots {\cal O}_n
\left(M_h \frac{d}{d M_h} \frac{1}{g_h^2 (M_h)}\right)
    \frac{1}{16} \int d^4 x d^2 \theta W W \rangle
    =
    -\langle {\cal O}_1 \cdots {\cal O}_n
    \int d^4x d^2\theta M_h \mbox{Tr}\phi^i_h
    \phi^i_h \rangle} \nonumber \\
  & & = -\langle {\cal O}_1 \cdots {\cal O}_n \left(
  \frac{1}{16} \int d^4 x d^2 \theta \frac{-3 t_2 (A)}{8\pi^2} W W
  + \mbox{higher-dimensional $D$-terms of } O(1/M^4) \right) \rangle ,
\nonumber \\
\label{eq:ME}
\end{eqnarray}
where we dropped operators such as $\mbox{Tr} \bar{\phi} e^{2V} \phi
e^{-2V}$ in $S_0$ which only produce contributions suppressed by $M_h$.  In
fact, the $F$-term $WW$ in the above equality can be shown to be exact
using the instanton argument given in the appendix.  The
higher-dimensional $D$-terms can be dropped without modifying relevant
couplings in the bare Lagrangian as discussed in Sec. 2.  By combining
Eqs.~(\ref{eq:change1}) and (\ref{eq:ME}), we find that the low-energy
physics can be kept fixed by changing the holomorphic gauge coupling
according to the one-loop law
\begin{equation}
  M_h \frac{d}{dM_h}
  \frac{1}{g^2_h (M_h)} = \frac{3 t_2 (A)}{8\pi^2}
\end{equation}
when one changes $M_h$.  This explicit calculation verifies the exact
one-loop law for the change of holomorphic coupling derived indirectly from the
argument based on holomorphy given in Sec. 2: given 
a regularization preserving
holomorphy, the running of $1/g^2_h$ is {\it guaranteed} to be exhausted at
1-loop.

Note that it is only the mass term which breaks both the conformal symmetry
and the non-anomalous $U(1)_R$ symmetry under which all three
$\phi^{i}$ have charge $2/3$.  Therefore the
response under dilation and $U(1)_R$ transformations are described the same
matrix element. This is the anomaly multiplet structure:
\begin{equation}
        \frac{1}{2} \left(\theta^{\mu}_{\mu} + i 
        \frac{3}{2} \partial_{\mu} j_{R}^{\mu} \right)
        = \int d^{2} \theta M_h
        \mbox{Tr}(\phi_{h}^{i}  
        \phi_{h}^{i}),
\end{equation}
where the right hand side can be replaced by $\frac{1}{16} \int d^2
\theta \frac{-3 t_2(A)}{8\pi^2} W W$ in the zero-momentum limit as in
Eq. (6.5).  These are indeed the correct trace and $U(1)_R$ anomalies for
$N=1$ pure SUSY YM in holomorphic normalization. 

Now that we have defined what we mean by $1/g^2_h(M),1/g^2_c(M)$, 
we can relate them to each other. We want to make the change of variable 
$\phi_{h}^i = g_c \phi_{c}^i,V_h=g_c V_c$. From the previous section, 
we know that the Jacobian of $V$ 
cancels the one from one adjoint multiplet, but this
leaves the Jacobian for 2 adjoint multiplets 
left uncancelled. Therefore, 
\begin{eqnarray}
\lefteqn{
  {\cal D}(g_c V_c)\prod_{i=1}^3 {\cal D}(g_c \phi_{c}^i) = {\cal D}V_c
  \prod_{i=1}^3 {\cal D} \phi_{c}^i }\nonumber \\ 
&\times&\exp\left(-2\times \frac{1}{16}\int d^4 y \int d^2 \theta
  \frac{2 t_2(A)}{8 \pi^2} \mbox{ln}g_c W^a(g_c V_c) W^a(g_c V_c) +
  \mbox{h.c.} + \cdots \right)
\end{eqnarray}
where $\cdots$ refers to the extra terms needed to make the Jacobian 
$N=4$ supersymmetric (see appendix A.5). Note that there are
no higher dimension $D$ terms in the above Jacobian: since the $N=4$ theory
is finite, the cutoff used in computing the Jacobian can be taken to infinity
with no difficulty. Therefore, we do not need to use a
non-renormalisation argument to justify dropping the $D$ terms in the Jacobian,
as they are simply absent.

Using this result, we find
\begin{eqnarray}
  \lefteqn{
    Z=\int {\cal D}V_h\prod_i{\cal D}\phi_{h}^i }\nonumber \\
  &&\times\exp\left(-\frac{1}{16}\int d^4y \int d^2 \theta
    \frac{1}{g^2_h(M_h)} W^a(V_h)W^a(V_h)+ M_h \mbox{Tr}(
    \phi_{h}^i \phi_{h}^i) + \mbox{h.c.} + \cdots\right) \nonumber \\
  && =\int {\cal D}(g_c V_c)\prod_i{\cal D}(g_c \phi_{c}^i) \nonumber \\ 
  &&\times\exp\left(-\frac{1}{16}\int d^4 y \int d^2 \theta
    \frac{1}{g^2_h(M_h)} W^a(g_c V_c)W^a(g_c V_c) + g_c^2 M_h \mbox{Tr}
    (\phi_{c}^i \phi_{c}^i) + \mbox{h.c.} + \cdots \right) \nonumber \\
  && =\int {\cal D}V_c\prod_i{\cal D}\phi_{c}^i \nonumber \\
  &&\times\exp\left(-\frac{1}{16}\int d^4y \int d^2
    \theta \left(\frac{1}{g^2_h(M_h)} + \frac{4 t_2(A)}{8 \pi^2}
      \mbox{ln}g_c\right) W^a(g_c V_c) W^a(g_c V_c) \right. \nonumber \\
  && \phantom{\times \exp} \left.
    + g_c^2 M_h \mbox{Tr}(
    \phi_{c}^i \phi_{c}^i) + \mbox{h.c.} + \cdots 
    \vphantom{\frac{1}{g^2_h(M_h)}}
  \right). 
\end{eqnarray}
Defining $M_c = M_h g_c^2$, we must have 
\begin{equation}
  \frac{1}{g_c^2(M_c)} = Re\left(\frac{1}{g_h^2(M_h=M_c/g_c^2)}\right)
  + \frac{4 t_2(A)}
  {8 \pi^2} \mbox{ln}g_c(M_c).
\end{equation}
Using the 1-loop law for $1/g^2_h$, 
\begin{equation}
  1/g^2_h(M_c/g_c^2) =
  1/g^2_h(M_c) - \frac{3 t_2(A)}{8\pi^2} \mbox{ln} g_c^2(M_c),
\end{equation}
we finally have
\begin{equation}
  \frac{1}{g_c^2(M_c)}= Re \left(\frac{1}{g^2_h(M_c)}\right) 
  - \frac{2 t_2(A)}{8 \pi^2} \mbox{ln} g_c(M_c)
\end{equation}
which is precisely the Shifman-Vainshtein formula (\ref{svformula}).
Because the holomorphic coupling has already been shown to run only at
one-loop, the canonical coupling in this explicit regularization
follows the NSVZ $\beta$-function.

One can repeat exactly the same exercise using finite $N=2$ theories.
In $N=1$ language, these theories contain the vector multiplet $V$ and
a chiral multiplet $\phi$ in the adjoint representation forming the
pure $N=2$ vector multiplet, as well as vector-like pairs of chiral
fields $Q_i,\tilde{Q}_i$, chosen so the 1-loop $\beta$ function
vanishes
\begin{equation}
  b_0 = 3 t_2(A) - t_2(A) - \sum_i t_2(i) = 0.
  \label{N2v}
\end{equation}
Suppose we wish to regulate an $N=1$ theory with the multiplets
$Q_i,\tilde{Q}_i$ (an example would be SUSY QCD with $2N$ flavors).
This can be done by starting with the $N=2$ theory with a mass term
$M_h \mbox{Tr} \phi_h^2$ added to the adjoint, which preserves the
finiteness of the theory \cite{softN2}.  If we now wish to go back to
canonical normalization for the gauge kinetic terms, we make the
change $V_h=g_c V_c, \phi_h = g_c \phi_c$. As mentioned previously, in
this case contributions from the vector and chiral multiplets cancel
in the Jacobian. However, the
mass term for the adjoint becomes $M_c \phi_c^2= M_h g_c^2 \phi_c^2$,
and so
\begin{equation}
  \frac{1}{g_c^2(M_c)} = Re\left(\frac{1}{g^2_h(M_h=M_c/g_c^2)}\right).
\end{equation} 
Using the 1-loop law for $1/g^2_h$ together with $3 t_2(A) - \sum_i
t_2(i)=t_2(A)$, we again find the Shifman-Vainshtein formula
\begin{eqnarray}
  \frac{1}{g_c^2(M_c)}=Re\left(\frac{1}{g^2_h(M_c/g_c^2)}\right) &=& 
  Re \left(\frac{1}{g^2_h(M_c)}\right) - 
  \frac{3 t_2(A) - \sum_i t_2(i)}{8 \pi^2}
  \mbox{ln} g_c^2(M_c) \nonumber \\ 
  &=&Re \left(\frac{1}{g^2_h(M_c)}\right) 
  - \frac{2 t_2(A)}{8 \pi^2} \mbox{ln} g_c(M_c).
\end{eqnarray}

In the case where the hypermultiplets are $2N$ flavors of an $SU(N)$
gauge group, we can add mass terms to some of the hypermultiplets,
thereby regularizing an arbitrary $N=1$ $SU(N)$ with $N_f<2N$ flavors.
Similar considerations lead to the Shifman-Vainshtein relation
and the NSVZ $\beta$ function in this case as well.

We obviously cannot extend the regularization methods discussed in
this section to general $N=1$ theories, especially chiral ones.
However, our observation that an explicit regularization preserving
holomorphy and yielding NSVZ $\beta$ function exists for a class of
$N=1$ theories does support our argument.  One can hope that a
certain regularization is possible for general $N=1$ theories which
preserves manifest holomorphy, perhaps by higher-derivative
regularization for the vector multiplet \cite{Krivoshchekov} and the
infinite tower of Pauli--Villars regulators for chiral multiplets
\cite{Slavnov}.

It is noteworthy that the exact NSVZ $\beta$-function can be checked
by explicit perturbative calculations, since we have given an explicit
regularization scheme for $N=1$ SUSY YM.  The procedure will be as
follows.  One works out certain Green's functions ({\it e.g.}\/ gauge
field two-point function in background field method) as a function of
external momenta, bare coupling and the regulator mass $M$.  Then one
tries to change the cutoff and the bare coupling at the same time to
keep the Green's function fixed.  In this way, the correct
$\beta$-function for the Wilsonian coupling constant can be
determined.  Since the theory is finite, there should be no ambiguity
in the analysis as long as the Green's function under study is free
from infrared singularities.

\section{Conclusions}
\setcounter{footnote}{0}
\setcounter{equation}{0}

In this paper, we hope to have clarified some of the mysteries
surrounding the gauge coupling $\beta$ functions for SUSY gauge
theories. The result is quite simple: if we work with the holomorphic
bare Lagrangian with a cutoff $M$, the change in $1/g^2_h$ needed to
keep the low energy physics fixed as the cutoff is changed from $M$ to
$M'$ is exhausted at 1-loop. However, since the rescaling of the
vector multiplet in going to canonical normalization for the matter
fields is anomalous, the gauge coupling $g_c$ in the theory with
canonical kinetic terms is not equal to $g_h$. The $F$ terms in this
Jacobian can be determined exactly, while the $D$ terms are suppressed
by powers of the cutoff. In a non-supersymmetric theory, these higher
dimension operators can in general feed back in to the coefficient of
relevant operators at higher orders; however in our case the higher
dimension $D$ terms are forbidden from doing so by the
non-renormalization theorem.  The final relationship between $g_c$ and
$g_h$ is given by the Shifman-Vainshtein formula (\ref{svformula}),
and the change in $g_c$ upon moving the cutoff from $M$ to $M'$ is
given by the NSVZ $\beta$ function (\ref{NSVZbeta}).  Our analysis
does not encounter any 
subtleties from infrared physics because we never refer to 1PI
effective actions.  All the discussions are on bare couplings within
the framework of Wilsonian effective action with a regularization in
the ultraviolet.  This is desirable because we can make a separation
between the ultraviolet structure of the theories (which determine the
evolution of the couplings) and model-dependent, dynamical effects
from infrared singularities.  We have understood that $N=2$ theories
have only one-loop $\beta$-function because the rescaling anomaly
cancels between the vector multiplet and adjoint chiral multiplet in
$N=1$ language. Finally, we have shown that certain $N=1$ theories can
be regularized with a cutoff $M$ starting from finite $N=4$ and $N=2$
theories, in a way preserving manifest holomorphy.  In these theories,
we have demonstrated that the Shifman-Vainshtein relation, and hence
the NSVZ $\beta$-function, holds. The claimed exactness of the $\beta$
function can then at least in principle be checked by direct
calculation in these explicitly regularized theories.

\section*{Acknowledgements}

We thank Dan Freedman, Bogdan Morariu, Hirosi Ooguri, and Bruno Zumino
for many useful discussions and comments on the manuscript.  NAH also
thanks Hsin-Chia Cheng and Takeo Moroi for discussions.  This work was
supported in part by the Director, Office of Energy Research, Office
of High Energy and Nuclear Physics, Division of High Energy Physics of
the U.S. Department of Energy under Contract DE-AC03-76SF00098 and in
part by the National Science Foundation under grant PHY-90-21139.  NAH
was also supported by NSERC, and HM by Alfred P. Sloan Foundation.

\appendix

\section{Rescaling Anomaly}
\setcounter{footnote}{0}
\setcounter{equation}{0}

\subsection{Generalities}
\label{subsec:generalities}

In this appendix, we discuss various aspects of the anomaly incurred
in changing the normalization of fields in the path integral, which we
call the rescaling anomaly. The rescaling of a quantum field is simply
a change of variables $\phi(x) = e^{\alpha} \phi'(x)$. In a general
non-supersymmetric theory, this change of variable is not unitary and
is not expected to leave the measure invariant; the Jacobian for this
transformation has power ultraviolet divergences and is highly
regularization dependent. Nevertheless, once a specific choice of
regularization is made, one must carefully account for the correct
Jacobian when making rescalings in the path integral.

In a supersymmetric theory, the situation is different: the
transformation $f(x,\theta,\bar{\theta})= e^{\alpha}
f'(x,\theta,\bar{\theta})$ for superfields {\it does} naively leave
the measure ${\cal D} f$ invariant, since the Jacobians from bosonic
and fermionic components naively cancel.  Of course, whether or not a
non-trivial Jacobian exists depends both on what type of
regularization is used and the symmetries which need to be preserved.
In the case of supersymmetric gauge theories, the preservation of
gauge invariance and supersymmetry force a non-trivial Jacobian for
the rescaling of both chiral and vector multiplets. The calculation of
these Jacobians is the main purpose of this appendix. First, however,
some preliminary remarks are in order.

We usually do not encounter the rescaling anomaly in perturbation
theory.  The reason is somewhat trivial: by convention, we employ
canonical normalization for bare fields and never change the
normalization.  The wave function renormalization is applied to the
fields in the 1PI effective action, where the rescaling of the fields
is nothing more than a relabeling of variables, as the 1PI effective
action is a classical object and no further functional integration is
done. On the other hand, a Wilsonian action ({\it i.e.}\/ a bare
theory defined with its cutoff) retains quantum fields beneath the
cutoff, which are then integrated over.  If one rescales the fields in
a Wilsonian effective action, the correct Jacobians must be taken into
account. The Jacobian gives the modification of the bare Lagrangian
necessary to keep the physics fixed after rescaling the quantum
fields, at a fixed value of the cutoff. In principle, with a given
cutoff, we just need to compute in the theories before and after the
rescaling, and explicitly see what changes are necessary in the bare
Lagrangian in order to keep all amplitudes fixed.

In practice, however, when we compute Jacobians, they are typically
regularized by hand in a way that preserves the important symmetries.
It is not a priori clear how the regularization of Jacobian is related
to the way in which the full theory is regularized; in principle, any
given regularization of the full theory should specify the
regularization of the Jacobians. In some cases, there is no problem
with being sloppy about this point. For example, in the case of the
chiral anomaly, we know that the Jacobian is completely topological in
nature and is independent of the way in which the theory is
regularized (providing the regularization is gauge-invariant).
Therefore, directly regularizing the Jacobian of a chiral
transformation, say by Gaussian damping as in the Fujikawa method
\cite{Fujikawa-chiral}, will give us the exact answer ({\it i.e.}\/ it
gives us the exact modification of the bare Lagrangian needed to keep
the physics fixed), since the answer is regulator independent. In
other cases, however, we have to be more careful. For instance, in the
case of the Jacobian for dilation \cite{Fujikawa-Weyl}, the regulator
independent pieces correctly reproduce the 1-loop $\beta$ function,
but where do all the higher order contributions to the $\beta$
function come from? The answer must be that either the higher
dimension operators in the Jacobian can not simply be thrown out, or
the quartic divergence in the Jacobian contains hidden dependence on
the fields at higher orders.  Recall that the higher-dimension
operators can only be set to zero after an appropriate modification of
the relevant couplings, presumably providing the higher order
corrections \cite{Wilson}. In this case, while we can 
regularize the Jacobian to get the 1-loop $\beta$ function, the way in
which the Jacobian is regularized must be {\it derived} from the
regularization of the full theory in order to get the higher order
corrections.

Having said this, in this appendix we will regularize all the
Jacobians we encounter by hand.  The reason is that, somewhat
analogous to the situation with the chiral anomaly, we can be sloppy
about the regularization here, for the following two reasons.  First,
the Jacobians come out automatically finite, and there is no concern
about an infinite constant changing the result at higher orders.
Second, the $F$ term in the Jacobian can be exactly computed and is
regularization independent. Third, while the Jacobian (regulated by
hand) does contain $D$ terms suppressed by powers of the cutoff, we
don't need to know the precise way in which this is related to how the
full theory is regularized; the usual supersymmetric
non-renormalization theorem makes it impossible for these $D$ terms to
ever feed back into an $F$ term like $WW$, and so they are truly
irrelevant for our interests.

Another remark is that the dimensional regularization does not produce
a rescaling anomaly.  This is a consequence of the following identity:
$\int d^D p \mbox{ const} = 0$.  When one employs dimensional
regularization (or more correctly, regularization by dimensional
reduction), the would-be effect of the rescaling anomaly appears as a
part of conventional perturbation theory.  Indeed, in perturbation
theory using dimensional regularization, the two-loop contribution to
the $\beta$-function, which we describe as a consequence of the
rescaling anomaly, appears from infrared uncertain terms $\sim 0/0$
\cite{0/0}.  However, this does not necessarily imply that the
two-loop contribution is coming from the infrared, since
dimensional regularization mixes up infrared and ultraviolet effects.

\subsection{Supersymmetric Path Integrals}
\label{subsec:super}
Since the vacuum energy vanishes in a supersymmetric background, the
path integral around a supersymmetric background is simply unity, and
we do not expect any anomalous rescaling Jacobian in this case.  Let
us first see how this works for for an $N=1$ chiral supermultiplet.
The Lagrangian is, given in terms of components,\footnote{This is a
  Euclidean Lagrangian \cite{Nicolai}.  There is no distinction
  between upper and lower indices.  The spinors $\psi$ and
  $\bar{\psi}$ are not related by complex conjugation; they must be
  treated as completely independent.  The auxiliary fields $F$ and
  $\bar{F}$ are also independent, and in fact, it is necessary to
  rotate their contours from $F$ to $iF$, $\bar{F}$ to $i\bar{F}$ to
  make the Gaussian integral over $F$, $\bar{F}$ fields possible.  We
  write Lagrangians before the rotation of $F$, $\bar{F}$ fields so
  that the correspondence to the Minkowski Lagrangian is more clear}
\begin{equation}
  {\cal L} = \int d^4 x e^{-2\alpha} \left( \partial_\mu \bar{\phi}
    \partial^\mu \phi + \bar{\psi}  {\not\!\partial} \psi - \bar{F}
    F\right)
  + \int d^4 x e^{-2\alpha} m \left( \frac{1}{2} \psi \psi + \phi F
    \right) + \mbox{h.c.}
\end{equation}
where we have for convenience included the $e^{-\alpha}$ factor in the
Lagrangian. Naively, if we just redefine $\phi = e^\alpha \phi'$, and
if the measure is invariant, nothing should depend on $\alpha$.  In
this trivial free theory, this is certainly the case.  The path
integral is given by
\begin{equation}
  Z = \frac{\mbox{det}(e^{-2\alpha}(\!{\not\!\partial} + m))}
  {\mbox{det}(e^{-2\alpha}(-\Box+m^2)) \mbox{det}(e^{-2\alpha})} = 1.
\end{equation}
The determinant in the numerator is taken over two-dimensional spinor
space, and hence $e^{-2\alpha}$ factor is counted twice.  Not
surprisingly, the $\alpha$ dependence drops out only when 
the auxiliary component is included; one cannot simply replace the
$F$-component 
by its solution to the equation of motion.  It is important to keep
the $N=1$ off-shell multiplet structure in path integrals.\footnote{We
  fortunately do not need off-shell multiplets of extended
  supersymmetry.}

In fact, the Jacobian for rescaling a chiral superfield can be
calculated directly \`a la Fujikawa without referring to the
determinants.  The Jacobians can be regularized by the kinetic
operator.  For a massive chiral superfield, one can use the Gaussian
regularization 
\begin{equation}
  e^{-t (-L+m^2)},
\end{equation}
where $t=1/M^2$ is an ultraviolet cutoff, and $L = \bar{D}^2 D^2/16 =
\Box$ when acting on a chiral superfield.  The Jacobian is trivial
because of a supersymmetric cancellation
\begin{equation}
  \ln J = \alpha \left(\mbox{Tr}_{\phi} e^{-t (-\Box + m^2)}
                - \mbox{Tr}_{\psi} e^{-t (-\Box +m^2)}
                + \mbox{Tr}_{F} e^{-t (-\Box + m^2)} \right) = 0.
\end{equation}
Again, the Jacobian from the $F$-component must be included, and in
this case cancels the contributions from the scalar and spinor
components.

\subsection{Chiral Multiplets}

Here we calculate the rescaling anomaly of a chiral multiplet in
background gauge field.  This calculation was done first by Konishi
and Shizuya \cite{KS} using the superfield formalism.  We repeat their
analysis in terms of component fields in order to gain a better
intuition on the anomalous Jacobian.

The path integral of a chiral multiplet in a gauge field 
background is given by
\begin{equation}
        \int {\cal D}\phi {\cal D}\psi {\cal D}F
        {\cal D}\bar{\phi} {\cal D}\bar{\psi} {\cal D}\bar{F}
        e^{-\int d^{4}x \left( |D\phi|^{2} + \bar{\psi} {\not D}
        \psi - \bar{F} F\right)} .
\end{equation}
We will discuss the case of Abelian gauge theory with the covariant
derivative $D_{\mu} = \partial_{\mu} - i A_{\mu}$, but the extension
to non-abelian case is straight-forward.  We calculate the anomalous
Jacobian of the measure ${\cal D}\Phi = {\cal D}\phi {\cal D}\psi
{\cal D}F$ under the rescaling of the chiral superfield $\Phi(y,
\theta) = \phi(y) + \sqrt{2} \theta \psi(y) + \theta^{2} F(y)$.  Note
that we treat $\phi$ and $\bar{\phi}$ etc independently.

Under the rescaling $\Phi = e^{\alpha} \Phi'$, with $\alpha$ a general
complex number, we formally have
\begin{equation}
  {\cal D}\Phi = 
        {\cal D}\phi {\cal D}\psi {\cal D}F
        = {\cal D}\phi' (\mbox{det} e^{\alpha})
        {\cal D}\psi' (\mbox{det} e^{\alpha})^{-2}
        {\cal D}F' (\mbox{det} e^{\alpha}) = {\cal D}\Phi' J
\end{equation}
where all the Jacobian factors appear to cancel out.  However, we need
to regularize the Jacobians appropriately:
\begin{equation}
        \ln J =  \alpha \left( \mbox{Tr}_{\phi} e^{t (D_\mu)^{2}}
                - \mbox{Tr}_{\psi} e^{t {\not D}^{2}}
                + \mbox{Tr}_{F} e^{t (D_\mu)^2} \right)
\label{eq:Jchiral}
\end{equation}
and they may not cancel out exactly as we will see below.  Note that 
the contribution from $\psi$ is a trace over two-component spinor 
space.  The expression is proportional to $(1 - 2 + 1)=0$ 
with trivial background $D_{\mu} = \partial_{\mu}$, but 
gives a non-vanishing result for non-trivial background gauge fields.

The above choice of the Gaussian regularization is motivated by the
following reason.  In the case where $\alpha$ is imaginary, we have
chiral rotation on the fermion fields, and the Jacobian is the one
associated with the chiral anomaly, so the Fujikawa method suggests
the usual ${\not\!\! D}^{2}$ Gaussian damping.  Furthermore, if the
fermion fields are expanded in eigenmodes of ${\not\!\! D}^{2}$, their
kinetic term is diagonal and the symmetries of the action are
manifest.  This suggests that the scalar component be damped by its
kinetic term $D_{\mu}^2$.  Furthermore, we know that the anomalous
Jacobians in a trivial background $A_\mu = 0$ must cancel between different
components in the same supermultiplet because supersymmetry fixes the
normalization of path integrals to unity.  Therefore, we must choose a
Gaussian regularization for the Jacobian of the auxiliary component
which cancels 
the anomalous Jacobians from the scalar and spinor components in the
trivial background, and hence
it must be $e^{t(\partial_\mu)^2}$ in the trivial background.  The only
possible gauge covariant extension is $e^{t(D_\mu)^2}$ because the
auxiliary component transforms the same way as the other components
under the ordinary gauge transformations. We will see later that this
regularization of the components can be justified in a manifestly
supersymmetric way \cite{KS}, but the above arguments are a quick
route to the correct answer.

For the case of constant background field strength and charge $+1$
chiral superfield, the traces (heat kernels) can be evaluated
explicitly using standard harmonic oscillator methods.  For
completeness, we review the methods here.  (For a manifestly
supersymmetric calculation of the heat kernel for a chiral superfield,
see \cite{SY}).  We employ background gauge field with constant
electric and magnetic fields, because the cutoff dependence of the
result can be explicitly seen without approximations.  One can
diagonalize the field strength tensor $F_{\mu\nu}$ to the form
\begin{equation}
        F_{\mu\nu} = \left( \begin{array}{cccc}
                0& E& 0& 0\\ -E& 0& 0& 0\\
                0& 0& 0& B\\ 0& 0& -B& 0 \end{array} \right).
\end{equation}

Let us pick a gauge with $A_1 = -E x_2/2, A_2= E x_1/2,A_3=-B
x_4/2,A_4=B x_3/2$, which clearly reproduces the above $F_{\mu
  \nu}$.\footnote{In this gauge $x^{\mu} A_{\mu} = 0$. This will prove
  useful when we consider the dilation anomaly in Appendix B, since
  the generator of co-ordinate dilations $x^\mu \partial_\mu$ is gauge
  invariant: $x^\mu \partial_\mu = x^\mu D_\mu$.}  Then, with $p_\mu =
-i \partial_\mu$, we have
\begin{eqnarray}
  \mbox{Tr} e^{t (D_\mu)^2} &=& 
  \mbox{Tr} e^{-t \left((p_1 - \frac{E}{2}x_2)^2 + (p_2 + \frac{E}{2} x_1)^2 
      + (p_3 - \frac{B}{2}x_4)^2 + (p_4 + \frac{B}{2} x_3)^2\right)} 
  \nonumber \\
  &=& \mbox{Tr} e^{-t H(E)} \mbox{Tr} e^{-t H(B)}
\end{eqnarray}
where 
\begin{eqnarray} 
  H(E)&=&\left( p_1 - \frac{E}{2} x_2 \right)^2 
  + \left( p_2 + \frac{E}{2} x_1 \right)^2 \nonumber \\ 
  &=& p_1^2 + p_2^2 + \left(\frac{E}{2}\right)^2 (x_1^2 + x_2^2) 
  - E(p_1 x_2 - p_2 x_1) 
\end{eqnarray}
Defining the usual harmonic oscillator raising and lowering operators as 
\begin{equation}
  p_\mu = \sqrt{\frac{2}{E}}\left(\frac{a_\mu - a^{\dagger}_\mu}
    {\sqrt{2}i} \right), \; 
  x_\mu = \sqrt{\frac{E}{2}}\left(\frac{a_\mu + a^{\dagger}_\mu}{
\sqrt{2}}\right)
\end{equation}
and further defining $a_0 = (a_L + a_R)/\sqrt{2}, a_1 = (a_L -
a_R)/\sqrt{2} i$, we find $H(E) = E(2 a_L^{\dagger} a_L + 1)$. Then
\begin{equation} 
  \mbox{Tr} e^{-t H(E)} = \sum_{n_L,n_R} e^{-t E (2 n_L + 1)}.
\end{equation}
The apparently divergent sum over $n_R$ is just proportional to the
area of the $(x_1,x_2)$ space. To see this, suppose that $(x_1,x_2)$
space is confined within a circle of radius $L$. Then, we should only
sum over the harmonic oscillator modes where $\langle n_L,n_R|(x^2 +
y^2)|n_L,n_R \rangle = 2/E (n_L + n_R + 1) < L^2$. It is then trivial
to perform the sum in the above, and we find
\begin{eqnarray}
  \mbox{Tr} e^{-t H(E)} &=& \frac{L^2 E}{2} \left(
    \frac{e^{-t E}}{1 - e^{-2 t E}} + {\cal O}(\frac{1}{L^2 E})
    \right) 
    \nonumber \\
    &=& \frac{(\pi L^2) E}{4 \pi} \frac{1}{\sinh tE} = 
    \frac{1}{4 \pi} \int dx_0 dx_1 \frac{E}{\sinh tE}, 
\end{eqnarray}
where in the second line we drop all the subleading terms in the large
area limit.\footnote{This result can be obtained with no
  approximations if we consider the system on a torus.}  Finally, then
\begin{equation}
  \mbox{Tr} e^{t (D_\mu)^2} 
  = \frac{1}{16 \pi^2} \int d^4x EB \frac{1}{\sinh tE \sinh tB}.
\end{equation}
The extension to the other heat kernels we need are straightforward.
For instance, in the case of the Dirac operator,
\begin{equation}
  {\not\!\! D}^{2} = (D_\mu)^2 -\frac{i}{4} [\gamma^\mu,\gamma^\nu]F_{\mu \nu}
  = (D_\mu)^2 + \left( \begin{array}{cc}
      (E + B)\sigma^3 & 0\\0 & (-E + B)\sigma^3 \end{array} \right)
\end{equation} 
and so the heat kernel for left/right handed chiral fermions is 
\begin{eqnarray}
  \mbox{Tr}_{L,R} e^{t {\not D}^2} &=& \mbox{Tr}e^{t (D_\mu)^2}
  \times \mbox{Tr}e^{t(\pm E + B)\sigma^3} \nonumber \\
  &=& \frac{1}{16 \pi^2} \int d^4 x EB \frac{1}{\sinh tE \sinh tB}
\times 2 \mbox{cosh}t(E \pm B).
\end{eqnarray}
 
Having computed the heat kernels, from Eq.~(\ref{eq:Jchiral}) we obtain,
\begin{equation}
  \ln J = \alpha \left(\frac{1}{16\pi^{2}} \int d^{4}x E B 
    \frac{2 - 2 \cosh t(E + B)}{\sinh t E \sinh t B}
  \right) .
\end{equation}
Similarly the Jacobian from ${\cal D}\bar{\Phi}$ is
\begin{equation}
  \ln \bar{J} = \alpha^{*} 
  \left(\frac{1}{16\pi^{2}} \int d^{4}x E B 
    \frac{2 - 2 \cosh t(E - B)}{\sinh t E \sinh t B}
  \right) .
\end{equation}

The result is quite interesting in the following respects.  First of
all, it is free from ultraviolet divergences $t = M^{-2} \rightarrow
0$ because of the cancellation between bosonic and fermionic degrees
of freedom, and is well-defined.  Expanding the Jacobians in the
inverse power of cutoff, we have
\begin{equation}
  \ln J =  \alpha \frac{1}{16\pi^{2}} \int d^{4}x
  \left( - (E + B)^{2} + 
    \frac{(E^{2} - B^{2})^{2}}{12 M^{4}} 
    + O(M^{-8}) \right)
\end{equation}
In supersymmetric notation,
\begin{equation}
  \ln J = -\frac{1}{16} \int d^{2} \theta \frac{2 t_2(\Phi)}{8 \pi^2}
  \mbox{ln}(e^\alpha) W_{\alpha} W^{\alpha} + O(1/M^4), 
\end{equation}
while the higher order terms can be written as $D$-terms, $\int d^{4}
\theta (W_{\alpha} W^{\alpha}) (\bar{W}_{\dot{\beta}}
\bar{W}^{\dot{\beta}})/M^{4}$ etc.  In conventional analyses of
anomalies, one drops all terms suppressed by powers of cutoff.
However, one must keep all higher dimension operators in Wilsonian
effective actions with a finite ultraviolet cutoff.  This implies that
the rescaling anomaly is not one-loop exact.\footnote{In Wilsonian
  effective actions, the loop calculations are exhausted at one-loop
  when one integrates out an infinitesimal slice in the momentum space
  \cite{Wilson}.  However, the one-loop results produce higher
  dimension operators and they produce corrections to renormalizable
  operators when one contracts some of the fields in the higher
  dimension operators.  Following the same reasonings, the existence
  of higher dimension operators in the Jacobians suggests that there
  are higher loop effects. As we have argued in Sec.~2, 
  however, for
  supersymmetric theories, these higher dimension operators can never
  feed back into the coefficient of $F$ terms like $WW$, and so are
  not relevant to the running of the gauge coupling. They may modify
  renormalizable $D$ terms such as the kinetic term for the matter
  fields.}  We will see later, however, that the ``holomorphic'' part
of $J$ is actually one-loop exact.  Note also that the anomalous
Jacobians are non-trivial even for topologically trivial background
gauge fields, {\it e.g.}\/, $E\neq 0$ and $B=0$.

Second, it is useful to check the result with a pure imaginary $\alpha
= i\theta$ because it is then a phase change of the chiral superfield
$\Phi$ and the anomalous Jacobian reduces to that of the chiral
anomaly.  The Jacobian is
\begin{eqnarray}
        \ln J + \ln \bar{J} 
        &= & i \theta \left(\frac{1}{16\pi^{2}} \int d^{4}x E B 
                \frac{- 2 (\cosh t(E + B) - \cosh t(E-B))}
                        {\sinh t E \sinh t B}
                \right) \nonumber \\
        &= &i \theta \left( \frac{1}{16\pi^{2}} \int d^{4} x
                (-4) EB \right).
\end{eqnarray}
This is nothing but the second Chern class $F_{\mu\nu}
\tilde{F}^{\mu\nu} / 16\pi^{2} = 4 EB/16\pi^{2}$, and is indeed the
correct formula for the chiral anomaly.  It is $t$-independent and
does not depend on the precise manner in which the Jacobian is
regularized.  The reason behind the $t$-independence is its
topological nature; the Jacobian is actually an integer which
corresponds to the mismatch between the number of zero modes for
different chiralities.  It is believed that the Jacobian for the phase
rotation is exact for this reason.

Finally, $J$ simplifies drastically under an instanton background, $E 
= \pm B$.  If $E = - B$, the integrand vanishes and there is 
no anomaly (but $\ln \bar{J} \neq 0$).  On the other hand
if $E = + B$, the Jacobian becomes $t$-independent,
\begin{equation}
        \ln J = \alpha \left( \frac{1}{16\pi^{2}} \int d^{4}x E B
                (-4) \right) .
\end{equation}

The result under the instanton background can be understood in terms
of the zero modes, analogously to the case of the chiral anomaly.
First of all, an instanton background preserves half of the
supersymmetry.  Depending on $E=-B$ or $E=B$, either $W^{\alpha}$ or
$\overline{W}^{\dot{\alpha}}$ vanishes, and hence either $Q^\alpha$ or
$\bar{Q}^{\dot{\alpha}}$ supercharges are unbroken \cite{SV}.  Therefore, the
modes of the differential operators $(D_\mu)^2$ and ${\not\!\! D}^{2}$
have the same spectrum, and there is a cancellation of eigenvalues
between bosonic and fermionic determinants \cite{NSVZ}.  Let us see
this more explicitly.  The scalar field can be expanded in terms of
the eigenmodes of $(D_\mu)^2$ operator, $- (D_\mu)^2 \phi_{n} =
\lambda^{2}_{n} \phi_{n}$.  On the other hand, the squared Dirac
operator is given in the Weyl basis by
\begin{equation}
   ({\not\!\! D})^2 = (D_\mu)^2 - \frac{1}{2} \sigma^{\mu\nu} F_{\mu\nu} 
        = \left( \begin{array}{cc}
                (D_\mu)^2 - \sigma^{\mu\nu}F_{\mu\nu}/2 & 0 \\
                0 & (D_\mu)^2
                \end{array} \right),
\end{equation}
where we used the fact $\bar{\sigma}^{\mu\nu}F_{\mu\nu} = 
\bar{\sigma}^{\mu\nu} (F_{\mu\nu} - \tilde{F}_{\mu\nu})/2 = 0$ for an 
instanton.  Therefore, there are two eigenmodes
\begin{equation}
        \bar{\psi}_{n}^{1} =
        \left( \begin{array}{c} 0\\ 0 \\ \phi_{n} \\ 0 \end{array} \right),
        \qquad
        \bar{\psi}_{n}^{2} =
        \left( \begin{array}{c} 0\\ 0\\ 0 \\ \phi_{n} \end{array} \right), 
\end{equation}
with the eigenvalue $-({\not\!\!D})^2 = -(D_\mu)^2 = \lambda_{n}^{2}$ for 
this chirality.  The eigenmodes of $({\not\!\!D})^2$ with the opposite 
chirality are given by
\begin{equation}
        \psi_{n}^{1} = 
        \frac{1}{\lambda_{n}} i{\not\!\!D} \bar{\psi}_{n}^{1}, 
        \qquad
        \psi_{n}^{2} = 
        \frac{1}{\lambda_{n}} i{\not\!\!D} \bar{\psi}_{n}^{2}. 
\end{equation}
There are, however, zero modes of $(D_\mu)^2 - 
\sigma^{\mu\nu}F_{\mu\nu}/2$ which cannot be written in this form 
because they are not paired with the opposite chirality spinor.  We 
refer to them as $\psi_{0}^{i}$ with $i = 1, \cdots, n_{0}$, where 
$n_{0}$ is the number of zero modes.  Finally, $F$ can be expanded in 
the same eigenmodes of $(D_\mu)^2$ as the scalar component, $F_{n} = 
\phi_{n}$.  The path integral measure then reduces to the 
following form:
\begin{equation}
        {\cal D}\Phi = 
        \prod_{n}(d \phi_{n} d\psi_{n}^{1} d\psi_{n}^{2} dF_{n})  
        \prod_{i} d\psi_{0}^{i} .
\end{equation}
Under the rescaling $\Phi = e^{\alpha}\Phi'$, the Jacobians from
$\phi_{n}$, $\psi_{n}^{1,2}$ and $F_{n}$ precisely cancel: $d \phi_{n}
d\psi_{n}^{1} d\psi_{n}^{2} dF_{n} = d \phi'_{n} d\psi'^{1}_{n}
d\psi'^{2}_{n} dF'_{n}$.  However the Jacobians from the zero modes
remain: $\prod_{i} d\psi_{0}^{i} = e^{-n_{0} \alpha}\prod_{i}
d\psi'^{i}_{0} $.  This is why the anomalous Jacobian is given by the
second Chern class; it is the number of zero modes due to the index
theorem.\footnote{Of course, the Atiyah--Singer index theorem tells us
  only the difference in the number of zero modes between two
  chiralities which is a topological invariant.  For certain
  configurations, there may be extra accidental zero modes with equal
  numbers for both chiralities.  In this case, the same zero modes
  appear also for $\phi$ and $F$ at the same time, and the accidental
  zero modes do not contribute to the anomalous Jacobian.}

The true anomalous Jacobian (that is, the correct change of the bare
Lagrangian after rescaling which keeps the physics fixed) is in
general a complicated function of the field strength $W_{\alpha}$ and
$\bar{W}_{\dot{\alpha}}$.  However, the instanton method shows that
the part of the Jacobian which is holomorphic in $W$ is {\it exact}\/
(similarly for anti-holomorphic part).  An instanton background has
only $W\neq 0$ with $\bar{W}=0$.  Therefore calculation in an
instanton background determines the holomorphic part of the anomalous
Jacobian, and just counts the number of zero modes in the background.
The part of the Jacobian depending only on $W$ is hence cutoff
independent, and is expected to be exact due to the same reasonings as
the chiral anomaly case.  Indeed, the result cannot be modified by the
higher order perturbative corrections because the half of the
supersymmetry left unbroken guarantees the cancellation of higher
order corrections \cite{NSVZ}.  Since the Jacobian in the instanton
background determines the holomorphic dependence on $W$, we learn that
the $F$-term in the Jacobian is exact for arbitrary background.

Even though we have used component calculations, we must mention that
the Gaussian regularization we used in this subsection can be made
manifestly supersymmetric, as was done originally in \cite{KS}:
\begin{equation}
  \ln J = \alpha \mbox{STr} \left( e^{t L} \frac{-\bar{D}^2}{4} \right),
\end{equation}
where the factor $-\bar{D}^2/4$ restricts the trace over the
superspace only to the chiral one, with\footnote{When we write $D^2$,
  it means either the square of the supercovariant derivative, or the
  square of the $D$-component in the vector multiplet.  We hope they
  can be easily distinguished according to the context.  The gauge
  covariant derivative is always written as $(D_\mu)^2$.}
\begin{equation}
  L = \frac{1}{16} \bar{D}^2 e^{-2V} D^2 e^{2V} .
\end{equation}
Indeed, the operator $L$ reduces to $(D_\mu)^2$ both on the scalar and
$F$-components, while it is ${\not\!\!D}^2$ on the spinor component.
We can heuristically understand how this $L$ operator can be arrived
at in a manifestly supersymmetric way. In a trivial background, we
found in the last subsection that an appropriate Gaussian cutoff was
provided by the operator $\bar{D}^2 D^2/16$ which reduces to $\Box$ on
chiral superfields. We are looking for a gauge-covariant extension of
this operator. Since we will be taking the trace over the chiral
space, our candidate operator $L$ should transform as $L \rightarrow
e^{i \Lambda} L e^{-i \Lambda}$ under gauge transformation. This is
clearly satisfied by $16 L=\bar{D}^2 e^{-2 V} D^2 e^{2 V}$; under the
gauge transformation $e^{2V} \rightarrow e^{i \bar{\Lambda}} e^{2 V}
e^{- i \Lambda}$, and
\begin{equation}
  \bar{D}^2 e^{-2 V} D^2 e^{2V} \rightarrow \bar{D}^2 e^{i \Lambda} e^{-2V}  
  e^{-i \bar{\Lambda}} D^2 e^{i \bar{\Lambda}} e^{2V} e^{-i \Lambda}=
  e^{i \Lambda}\left(\bar{D}^2 e^{-2V} D^2 e^{2V}\right)e^{-i \Lambda}.
\end{equation}
We can motivate this operator in another way. In the component
analysis, the operator in the Gaussian damping was related to
operators appearing in the equations of motion, so we can try to get a
hint for the form of a manifestly supersymmetric operator by looking
at the supersymmetric equations of motion, which are $D^2 e^{2V}
\phi=0$. Of course, we can not use the operator $D^2 e^{2V}$ directly
in damping the chiral Jacobians, since it maps chiral fields to
anti-chiral ones. However, we can get an operator mapping chiral to
chiral fields by acting on the left with a $\bar{D}^2$ appropriately;
as we have seen we need to use $\bar{D}^2 e^{-2V}$ acting on the left
to insure gauge covariance of the operator. These heuristic tools for
finding manifestly supersymmetric regulators will prove useful in
subsection A.5, where we examine the $N=2$ structure of the anomalies
induced in rescaling hypermultiplets in $N=2$ theories.
 
One further check which can be made with component calculations is to
look at the rescaling Jacobian when there is a constant $D$ term
background, and compare to the the case with $E,B$ background; they
should combine appropriately into $\int d^2\theta WW$.  Decomposing
the $L$ operator above in the $D$-term background, we find $L=\Box -
D$ on the scalar, $L=\Box$ for on the spinor and $L=\Box+D$ on the
$F$-component of the chiral superfield. The anomalous Jacobian under
the rescaling is
\begin{equation}
  \ln J = -\alpha \left( \mbox{Tr}_{\phi} e^{-t (-\Box+D)}
                - \mbox{Tr}_{\psi} e^{-t \Box}
                + \mbox{Tr}_{F} e^{-t (-\Box - D)} \right)
              = -\alpha \frac{1}{16\pi^2} D^2 .
\end{equation}
This contribution is exactly what one expects from $\int d^2 \theta W
W$ operator with the same normalization as for the case of the $E,B$
background.

\subsection{Vector Multiplets}

In this subsection, we calculate the anomalous Jacobian of vector
multiplets under rescaling.  The basic idea in the calculation is the
following.  At a given configuration of the gauge field in the
functional space, the path integral measure is defined by the top form
on the cotangent space, modulo the directions of gauge degrees of
freedom.  The calculation of the Jacobian requires only the knowledge
on local properties around each point in the functional
space.  Therefore, we define the measure in terms of local
``fluctuations'' around a particular ``background'' configuration and
regularize it in terms of a ``background-gauge invariant'' operator.
Note that we are not employing a background gauge field in the sense
of background field formalism where the background is an external
classical field.  The ``background'' configuration in the calculation
is what needs to be integrated over when the full functional integral
is done.  However all the steps of the calculation strongly resembles
the background field formalism.

As emphasized in subsection \ref{subsec:super}, it is important to
retain the structure of off-shell multiplet in order to retain the
supersymmetric cancellation; in Wess-Zumino gauge we need the gauge
field $V_{\mu}$, gaugino $\lambda$ and the auxiliary field $D$.  Three
(transverse) components of $A_{\mu}$ and $D$ balance the four
components of $\lambda$, $\bar{\lambda}$ off-shell.  As discussed in
Sec.~2, we are interested in the anomalous Jacobian when one rescales
the vector multiplet to bring its kinetic term to the canonical
normalization.

We will work with the vector multiplet in the Wess--Zumino gauge,
$(V_\mu, \lambda, D)$.  The anomalous Jacobians from the path integral
measures of $\lambda$, $\bar{\lambda}$ and $D$ can be readily
calculated using the formulae presented in the previous subsection.
The discussion of the vector field requires care, since it is only the
transverse components which are included in the off-shell multiplet.
One can go through supersymmetric gauge fixing; it however requires
three Faddeev--Popov chiral superfields and many unphysical auxiliary
components with higher derivative kinetic terms.  We find it more
intuitive to work within the Wess--Zumino gauge with appropriate
projection on the transverse components.  We work on the vector field
with the background field formalism, and discuss the Jacobian from the
path integral measure of the ``quantum'' vector field $V^{\mu}$.  We
will come back to a manifestly supersymmetric method later.

Let us go through the conventional Faddeev--Popov procedure to reduce
the path integral volume of the quantum vector field only to its
transverse components, being careful to keep track of the
normalization of the path integral. As usual, we insert an identity
\begin{equation}
        1 = \int {\cal D}g \, \delta(D^{\mu} V^{g}_{\mu} - a)
                \mbox{det} (D^{\mu} D_{\mu})
\end{equation}
into the path integral, and rewrite the determinant factor using the
Faddeev--Popov ghost.  Here, $V^{g}$ is a gauge transformed vector
field according to the gauge function $g(x)$, and $D_\mu =
\partial_\mu - i A_\mu$ with respect to the background vector field
$A_\mu$.  The gauge group volume ${\cal D} g$ can be normalized to
unity, and it can be dropped from the path integral because the rest
of the integrand is gauge-invariant.  Care must be taken when
``smearing'' the gauge fixing condition $D^{\mu} V_{\mu} = a$ over the
arbitrary space-time dependent function $a(x)$.  To obtain the desired
gauge fixing term $\frac{\xi}{2 g^{2}} (D^{\mu} V_{\mu})^{2}$, one
must integrate over $a$ as
\begin{equation}
        e^{-\int d^{4} x \frac{\xi}{2g^{2}} (D^{\mu} V_{\mu})^{2}}
        = \frac{1}{N} \int {\cal D}a \, \delta(D^{\mu} V^{g}_{\mu} - a)
                e^{-\int d^{4} x \frac{\xi}{2g^{2}} a^{2}},
\end{equation}
where the normalization factor $N$ depends on the gauge coupling
constant, $N = (\mbox{det}(g^{2}/\xi))^{-1/2}$.  When one rescales the
gauge field from holomorphic to canonical normalization, this factor
$N$ also changes.  To keep track of this factor $N$, we write the path
integral over the gauge field $V^{\mu}$ as
\begin{equation}
  \frac{\int {\cal D}V {\cal D}c {\cal D}\bar{c}\,  
    e^{-S_{V}-S_{FP}-S_{gf}}}
  {\int {\cal D}a\,  e^{-\int d^{4} x \frac{\xi}{2g^{2}} a^{2}}}
\end{equation}
where $S_{V}$ is the action for the quantum gauge field $V_{\mu}$ in
the presence of a background, $S_{FP} = \int d^{4} x \bar{c} D_{\mu}
D^{\mu} c$ is the Faddeev--Popov term, and $S_{gf} = \int d^{4} x
\frac{\xi}{2g^{2}} (D_{\mu} V^{\mu})^{2}$ is the gauge fixing term.
Note that the Faddeev--Popov action does not have an overall $1/g^{2}$
and hence does not need to be rescaled.  On the other hand, both the
vector field $V^{\mu}$ and the ``smearing'' factor $a$ need to be
rescaled.  The kinetic operator for $V_{\mu}$ is given by
\begin{eqnarray}
  \lefteqn{
    (D_\mu)^2\delta_{\mu\nu} - D_{\nu} D_{\mu}
    -i \frac{1}{2} F_{\rho\sigma}(M^{\rho\sigma})_{\mu\nu}
    + \xi D_{\mu} D_{\nu} } \nonumber \\
  & &
  = (D_\mu)^2\delta_{\mu\nu} 
  -i F_{\rho\sigma}(M^{\rho\sigma})_{\mu\nu}
  +(\xi - 1) D_{\mu} D_{\nu}
\end{eqnarray}
where $M^{\rho\sigma}$ are SO(4) rotation generators.  The anomalous
Jacobian from rescaling $V^{\mu} = e^{\alpha} V'^{\mu}$ is given by
\begin{eqnarray}
  \lefteqn{
    \ln J_{V} = \alpha \mbox{Tr}_{V} e^{-t((D_\mu)^2\delta_{\mu\nu} 
      -i F_{\rho\sigma}(M^{\rho\sigma})_{\mu\nu}
      + (\xi - 1) D_{\mu} D_{\nu})}
    } \nonumber \\
  & &
  = \alpha \left(
    \mbox{Tr}_{V_{T}} e^{-t((D_\mu)^2\delta_{\mu\nu} 
      -i F_{\rho\sigma}(M^{\rho\sigma})_{\mu\nu})}
    + \mbox{Tr}_{V_{L}} e^{-t \xi D_{\mu} D^{\mu}} \right)
\end{eqnarray}
where we have decomposed the space into the transverse one
$V_{T}^{\mu} = V^{\mu} - \frac{D^{\mu} D^{\nu}}{(D_{\rho})^{2}}
V_{\nu}$ and the longitudinal $V_{L}^{\mu} = \frac{D^{\mu}
  D^{\nu}}{(D_{\rho})^{2}} V_{\nu}$.  The Jacobian from ${\cal D}a$ is
regularized uniquely as
\begin{equation}
  \ln J_{a} = \alpha \mbox{Tr}_{a} e^{-t\xi D_{\mu} D^{\mu}}
\end{equation}
to guarantee the cancellation of the Jacobian under a trivial
background, and the combination $\ln J_{V} - \ln J_{a}$ is independent
of the gauge parameter $\xi$.  Therefore, we can simplify the
calculation by taking $\xi=1$ (Feynman gauge).

The Jacobian of the $D$-component is regularized as
\begin{equation}
        \ln J_{D} = \alpha \mbox{Tr}_{D} e^{t (D_{\mu} D^{\mu})}
\end{equation}
which is the only one allowed by the gauge invariance.  Putting all 
factors together, the total Jacobian is given by
\begin{eqnarray}
  \lefteqn{
    \ln J = \alpha \left(
      \mbox{Tr}_{V} 
      e^{t ((D_\mu)^2\delta_{\mu\nu} 
        -i F_{\rho\sigma}(M^{\rho\sigma})_{\mu\nu})}
    \right.} \nonumber \\
  & & \left.
    - \mbox{Tr}_{a} e^{t (D_\mu)^2}
    - \mbox{Tr}_{\lambda} e^{t {\not D}^{2}}
    - \mbox{Tr}_{\bar{\lambda}} e^{t {\not D}^{2}}
    + \mbox{Tr}_{D} e^{t (D_{\mu} D^{\mu})}\right)
\end{eqnarray}
and the second term is the same as the last term.  We finally find
that the anomalous Jacobian is simply
\begin{equation}
  \ln J =  \alpha \left( \mbox{Tr}_{V} e^{t ((D_\mu)^2\delta_{\mu\nu} 
      -i F_{\rho\sigma}(M^{\rho\sigma})_{\mu\nu})}
    - \mbox{Tr}_{\lambda} e^{t {\not D}^{2}}
    - \mbox{Tr}_{\bar{\lambda}} e^{t {\not D}^{2}} \right) .
\end{equation}

To simplify the analysis, we take SU(2) gauge group, and take a
constant background field strength in $W^{3}$ gauge field.  $W^{+}$
carries a positive charge unity under the background.  The rest is the
calculation of the Jacobian from $W^{+}$ multiplet only.  The only new
heat kernel we need is
\begin{eqnarray}
  \mbox{Tr}_{V} e^{t ((D_\mu)^2\delta_{\mu\nu} 
    -i F_{\rho\sigma}(M^{\rho\sigma})_{\mu\nu})} &=&
  \mbox{Tr} e^{t(D_\mu)^2} \times \mbox{Tr} \, \exp\left(\begin{array}{cc} 
      2 i t E \sigma^2 & 0\\0 & 2 i t B \sigma^2 \end{array} \right) 
  \nonumber \\
  &=& \frac{1}{16\pi^{2}} \int d^{4}x
  E B \frac{2(\cosh 2tE + \cosh 2tB)}{\sinh t E \sinh t B} . 
\end{eqnarray}
The anomalous Jacobian of the vector multiplet ${\cal D}V = {\cal
  D}(e^{\alpha} V') = {\cal D} V' J$ is given by
\begin{equation}
  \ln J = -\alpha \frac{1}{16\pi^{2}} \int d^{4} x
  E B \frac{2(\cosh 2tE + \cosh 2tB) - 4 \cosh tE \cosh tB}
  {\sinh tE \sinh tB}
\end{equation}
As expected, there is no ultraviolet divergence $t = M^{-2}
\rightarrow 0$.  By expanding the expression in powers of $t$, one
finds
\begin{equation}
  \ln J = \alpha \frac{1}{16\pi^{2}} \int d^{4} x
  \left( 2 (E^{2} + B^{2}) + \frac{5}{6} 
    \frac{(E^{2} - B^{2})^{2}}{M^{4}} +
    O(M^{-8}) \right)
\end{equation}
The finite part is exactly opposite to the contribution of a chiral
superfield with the same charge.

As in the case of chiral multiplets, the Jacobian simplifies
drastically for an instanton background $E=B$, where it becomes
\begin{equation}
  \ln J = -\alpha \frac{1}{16\pi^{2}} \int d^{4} x
  E B \times 4
\end{equation}
and is cutoff independent.  This is again a consequence of the zero
modes.  We have discussed the zero modes of spinors already in the
previous subsection.  The eigenvalues of the operator $(D_\mu)^2 + i
M_{\rho\sigma} F^{\rho\sigma}$ are the same as the squared Dirac
operator except the zero modes.  It is useful to write a vector field
$V_{\mu}$ as a bi-spinor $V_{\alpha \dot{\alpha}}$ for this purpose.
The eigenmodes satisfy the equation
\begin{equation}
        - (D_\mu)^2 V
        + \sigma_{\rho\sigma} F^{\rho\sigma} V
        + V \bar{\sigma}_{\rho\sigma}^{T} 
                F^{\rho\sigma}
        = \lambda_{n}^{2} V .
\end{equation}
The point is that $\bar{\sigma}_{\rho\sigma} F^{\rho\sigma} =
\bar{\sigma}_{\rho\sigma}(F^{\rho\sigma} - \tilde{F}^{\rho\sigma})/2 =
0$ for the instanton background.  Therefore, the eigenequation becomes
exactly the same as that of spinors except a left-over free spinor
index $\dot{\alpha}$.  The eigenvalues of the vector multiplet are
exactly the same as those of the spinor $\lambda$ with twice as much
degeneracy.  Together with the analysis in the previous subsection, we
find the following spectrum.  For each non-zero eigenvalue of
$-(D_\mu)^2 = \lambda_{n}^{2}$, there are two modes for $\lambda$, two
modes for $\bar{\lambda}$, four modes for $V_{\mu}$.  However one of
the four modes for $V_{\mu}$ is longitudinal.  Since the longitudinal
mode is always accompanied by the corresponding mode in ${\cal D}a$
and the Jacobians cancel between them, we drop it from the discussion.
For $n_{0}$ zero modes for $\lambda$, there are $2n_{0}$ zero modes
for $V_{\mu}$, and all of them are transverse.  Therefore in the
instanton background, the path integral measure reduces to the
following:
\begin{equation}
        \int {\cal D}V = \int 
        \prod_{n} \left( d V_{n}^{1} d V_{n}^{2} d V_{n}^{3}
                d \lambda_{n}^{1} d \lambda_{n}^{2}
                d\bar{\lambda}_{n}^{1} d\bar{\lambda}_{n}^{2}
                d D_{n}\right)
        \prod_{i}^{2 n_{0}} d V_{0}^{i}
                \prod_{i}^{n_{0}} d\lambda_{0}^{i}
\end{equation}
When one rescales the whole vector multiplet, the contributions from
all non-zero modes cancel among themselves.  The anomaly under the
rescaling is therefore determined by $2 n_{0} - n_{0}$, which is the
opposite of the case of a chiral superfield.  Following the same
reasoning as in the previous subsection, the part of the Jacobian
holomorphic in $W$ is exact.

The final result of the Jacobian for a general non-abelian gauge group
is
\begin{equation}
  {\cal D}(g_c V_c) = {\cal D}(V_c)\, \exp\left(\frac{1}{16}\int d^4 y 
    \int d^2 \theta \frac
    {2 t_2(A)}{8 \pi^2} \, \mbox{ln}g_c \, W^a(g_c V_c) W^a(g_c V_c)+
    \mbox{h.c.} + {\cal O}(1/M^4)\right).
\end{equation}

A manifestly supersymmetric formulation of the Jacobian is possible in
the background field formalism \cite{GSR} but is technically more
complicated.  First of all, one needs three Faddeev--Popov ghosts $c$,
$c'$, and $b$, which are all chiral superfields.  The first two appear
in a rather conventional manner.  The delta functional for gauge
fixing $\delta(\bar{\cal D}^2 V - a) \delta({\cal D}^2 V - \bar{a})$
cannot be inserted to the path integral by itself because it varies
along the gauge volume.  Here and below, ${\cal D}_\alpha = e^{2 W^B}
\bar{D_\alpha} e^{-2 W^B}$ is the background-chiral supercovariant
derivative, with $e^{2 W^B} e^{2 \bar{W}^B} = e^{2 V^B}$ is the
background vector multiplet.  The gauge variation of the gauge field
$V$ is given by $e^{2 V'} = e^{i \bar{\Lambda}} e^{2 V} e^{-i
  \Lambda}$, or $2 \delta V = - iL_{V}[(\bar{\Lambda}+\Lambda) +
\coth(L_{V})(\Lambda - \bar{\Lambda})]$ for infinitesimal 
$\Lambda$, $\bar{\Lambda}$.  ($L_V$ acts as $L_V c
= [V, c]$ etc, and the formal expression $L_V \coth L_V$ is understood
in terms of its Taylor expansion.)  Therefore, one inserts the
combination
\begin{equation}
  \delta(\bar{\cal D}^2 V - a) \delta({\cal D}^2 V - \bar{a})
  \int {\cal D} c {\cal D} \bar{c} {\cal D} c' {\cal D}\bar{c}'
  e^{{\rm Tr} \int d^4 x d^4 \theta (\bar{c}'-c') L_{V}
  [(\bar{c}+c) + \coth(L_{V})(c-\bar{c})]}
\end{equation}
Note that the ghost fields $c$, $\bar{c}$ have the normalization of
the gauge parameters and hence do not have $1/g^2$ in front of the
Lagrangian.  The third one $b$ corresponds to the normalization factor
from $a$-integration in the component treatment.  It appears when one
``smears'' over the gauge fixing condition $\bar{{\cal D}}^2 V = a$
where $a$ is a chiral superfield.  To guarantee that the delta
functional is correctly replaced by a path integral without any
additional factors, one needs to compensate the integral over $a$ by
an integral over ghost $b$,
\begin{eqnarray}
  \lefteqn{
    \delta(\bar{\cal D}^2 V - a) \delta({\cal D}^2 V - \bar{a})} \nonumber \\
  & &
  \rightarrow \int {\cal D}a {\cal D}\bar{a} {\cal D}b {\cal D}\bar{b}
  \delta(\bar{\cal D}^2 V - a) \delta({\cal D}^2 V - \bar{a}) 
  e^{-\frac{1}{16 g^2} \int d^4 x d^4 \theta (\bar{a} a + \bar{b} b)}
  \nonumber \\ 
  & & = \int {\cal D} b {\cal D} \bar{b} e^{- \frac{1}{16 g^2}\int d^4 x
    d^4 \theta 
    ( ({\cal D}^2 V) (\bar{\cal D}^2 V) + \bar{b} b )}
\end{eqnarray}
One needs the normalization $1/g^2$ so that the gauge fixing term
after the $a$ integral combines with the gauge kinetic term in the
holomorphic normalization.  The $b$ integral can not be dropped
since $b$ is background-chiral, {\it i.e.}\/ it satisfies the
chirality condition $\bar{\cal D}_\alpha b = e^{2 W^B} \bar{D_\alpha}
e^{-2 W^B} b = 0$,\footnote{If necessary, one can rescale $b$-ghost to
  absorb the factor of 16 by properly changing the holomorphic gauge
  coupling constant $8\pi^2/g_h^2$ by $- C_A \ln 16$. } and 
  hence the path integral over $b,\bar{b}$
depends on the background gauge field.

The change from the holomorphic normalization to the canonical
normalization requires rescaling of the full vector multiplet and
$b$-ghost, but not $c$, $c'$ ghosts.  The Jacobian from the $b$-ghost
is the same as that from a chiral superfield in the adjoint
representation except with opposite sign.  The vector multiplet
produces Jacobians from all components, {\it i.e.}\/, $V = C +
i\theta\chi - i\bar{\theta}\bar{\chi} + i\theta^2(M+iN)/2
-i\bar{\theta}^2(M-iN)/2-\theta \sigma^\mu \bar{\theta} V_\mu +
i\theta^2 \bar{\theta}(\bar{\lambda} + i {\not\!\bar{\partial}}
\chi/2) -i \bar{\theta}^2 \theta(\lambda + i{\not\!\partial}
\bar{\chi}/2) + \theta^2 \bar{\theta}^2 (D/2 + \Box C/4)$.  The
Jacobian is regularized by the kinetic operator,
\begin{equation}
  \ln J = \alpha \mbox{STr}_V e^{t((D_\mu)^2 - {\cal
  W}^\alpha {\cal D}_\alpha + \bar{\cal W}^{\dot{\alpha}} \bar{\cal
  D}_{\dot{\alpha}})}
\end{equation}
which reduces to $(D_\mu)^2$ for $C$, $M$, $N$, $D$ components,
$(D_\mu)^2 + M_{\rho\sigma} F^{\rho\sigma}$ for the vector $V_\mu$,
and ${\not\!\!D}^{2}$ for $\chi$, $\lambda$.  Therefore, the addition
to the case in the Wess--Zumino gauge is $C$, $M$, $N$, $V_L$
(longitudinal component of $V_\mu$) and $\chi$, $\bar{\chi}$, and
hence is the same as an extra chiral superfield in the adjoint
representation.  This additional contribution is precisely canceled by
the Jacobian from the $b$-ghost and hence our component calculation is
justified from a manifestly supersymmetric framework.  Put another
way, the vector multiplet does not produce an anomalous Jacobian in a
manifestly supersymmetric analysis; this is because one needs two
powers of ${\cal D}_{\alpha}$ and two powers of $\bar{\cal
  D}_{\dot{\alpha}}$ to get a non-vanishing supertrace, and the
leading term is hence $WW\bar{W}\bar{W}$ which is a higher-dimensional
$D$-term.  The relevant Jacobian comes solely from the $b$-ghost;
therefore it is always the opposite of that from a chiral superfield
in the adjoint representation.

\subsection{$N=2$ invariance}

In the previous subsections, we calculated anomalous Jacobians of the
chiral and vector multiplet in $N=1$ supersymmetric gauge theories.  A
natural question is what happens when one studies theories with
extended supersymmetries.  Clearly, the Gaussian cutoff method can be
extended for the rescaling anomaly of hypermultiplets in $N=2$
theories.  An important question then is whether the rescaling of a
hypermultiplet produces both the $\int d^2 \theta WW$ operator and the
kinetic term of the adjoint superfield $\int d^4\theta 2 \mbox{Tr}
\bar{\Phi} e^{2 V}
\Phi e^{-2 V}$ needed to preserve $N=2$ supersymmetry.  Of course, the
hypermultiplets do not receive wave-function renormalization and its
rescaling is not necessary for the computation of $\beta$-function.
However, the rescaling of a hypermultiplet in the adjoint 
representation is necessary to derive the Shifman--Vainshtein formula
Eq.~(\ref{svformula})
from the $N=1$ SUSY YM regularized by $N=4$ theory in Sec.~6, and it 
is important to check that the Jacobian preserves extended 
supersymmetry.  The $N=4$ invariance of the Jacobian needed in Sec.~6 
follows trivially once the $N=2$ invariance is verified.

There is a superpotential coupling of a hypermultiplet $(Q,\tilde{Q})$
to the adjoint superfield $\Phi$ in the vector multiplet,
\begin{equation}
  \int d^2 \theta \sqrt{2} \tilde{Q} \Phi Q .
\end{equation}
To see that $\int d^4 \theta 2 \mbox{Tr} \bar{\Phi} e^{2V} \Phi
e^{-2V}$ is generated from rescaling the hypermultiplet, we need to
employ a background configuration of $\Phi$ such that the kinetic
operator does not vanish.  A convenient choice is when the
$F$-component of $\Phi$ does not vanish.  For simplicity, we discuss
$N=2$ supersymmetric QED, where $\Phi$ has only one component and is
electrically neutral.

Let us first find a manifestly N=2 supersymmetric Gaussian damping
operator.  In order to do this, we follow the strategy used in
subsection A.3 and use the supersymmetric equations of motion to infer
the form of the operator we need, this should work since the equations
of motion are certainly $N=2$ covariant.  The equations of motion are
\begin{equation}
  \left( \begin{array}{cc}
      D^2 e^{2V} & \sqrt{2} \bar{\Phi} \\ \sqrt{2} \Phi & \bar{D}^2 e^{-2V} 
    \end{array} \right)
  \left(\begin{array}{c}Q\\ \widetilde{Q}^{\dagger}\end{array}
  \right) = 0.
\end{equation}
As before, in order to find an operator that correctly maps (anti)
chiral to (anti) chiral fields, and which is moreover gauge covariant,
we form
\begin{eqnarray}
  L  \left(\begin{array}{c}Q\\ \widetilde{Q}^{\dagger}\end{array}
  \right)
  &=&\frac{1}{16}\left( \begin{array}{cc} \bar{D}^2 e^{-2V} & 0\\0 &
      D^2 e^{2V} \end{array} 
  \right) \left( 
    \begin{array}{cc} 
      D^2 e^{2V} & \sqrt{2} \bar{\Phi} \\ \sqrt{2} \Phi & \bar{D}^2 e^{-2V}
    \end{array} \right)\left(\begin{array}{c}Q\\
      \widetilde{Q}^{\dagger}\end{array} 
  \right) \nonumber \\
  &=& \frac{1}{16} \left( \begin{array}{cc} \bar{D}^2 e^{-2V} D^2
  e^{2V} & \sqrt{2} 
      \bar{D}^2
      e^{-V} \bar{\Phi} \\ \sqrt{2}D^2 e^{2V} \Phi & D^2 e^{2V} \bar{D}^2 
      e^{-2V} \end{array} \right) \left(\begin{array}{c}Q\\
  \widetilde{Q}^{\dagger}\end{array} 
\right).
\end{eqnarray}
For trivial gauge fields and a background $F$ component $F_{\Phi}$ for
$\Phi$, the action of $L$ on components is very simple: $\Box$ on the
fermion and $F$ components of $(Q,\tilde{Q})$, and
\begin{equation}
  L\left( \begin{array}{c} A_Q \\ \bar{A}_{\widetilde{Q}} \end{array} \right)
  = \left( \begin{array}{cc} \Box & \sqrt{2} \bar{F}_{\Phi} \\
  \sqrt{2} F_{\Phi} & 
  \Box \end{array} \right) \left( \begin{array}{c} A_Q \\
  \bar{A}_{\widetilde{Q}}  
\end{array} \right)
\end{equation}
on the $A$ components of $(Q,\widetilde{Q})$, 
so  $L$ has eigenvalues $\Box \pm \sqrt{2 F_{\Phi} \bar{F}_{\Phi}}$ on 
the space of $A$ components. The Jacobian is then
\begin{eqnarray}
  \ln J &=& \alpha \mbox{Tr}\left(e^{t (\Box + \sqrt{2 F_{\Phi}
        \bar{F}_{\Phi}})} +  
    e^{t (\Box - \sqrt{2 F_{\Phi} \bar{F}_{\Phi}})} - 2 e^{t \Box} \right)
  \nonumber \\   
  &=& \alpha
  \int d^4x \frac{d^4p}{(2 \pi)^4} \left( e^{-t(p^2 + \sqrt{2 F_{\Phi}
        \bar{F}_{\Phi}})} + 
    e^{-t(p^2 - \sqrt{2 F_{\Phi} \bar{F}_{\Phi}})} - 2 e^{-t p^2}
        \right) \nonumber \\ 
        &=& \alpha \frac{1}{4 \pi^2} \int d^4x F_{\Phi} \bar{F}_{\Phi}
        + O(1/M^4).
\end{eqnarray} 
This is nothing but the kinetic term of the adjoint superfield
$-\bar{F}_\Phi F_\Phi$ multiplied by $-\alpha/4\pi^2$.  On the other
hand the corresponding Jacobian in a gauge field background is
\begin{equation}
  \ln J = \alpha \frac{-1}{16\pi^2} \int d^4 x 
  \left(  (E+B)^2 + (E-B)^2 \right) + O(1/M^4)
  = \alpha \frac{-1}{4\pi^2} \int d^4 x
  \left( \frac{1}{2} (E^2 + B^2) \right) + O(1/M^4),
\end{equation}
which is again the gauge kinetic term multiplied by $-\alpha/4\pi^2$.
Therefore the anomalous Jacobian which we calculated comes out $N=2$
supersymmetric automatically.

\section{Trace Anomalies}
\setcounter{footnote}{0}
\setcounter{equation}{0}

In this appendix, we employ the same formalism as in the previous
appendix to work out the trace anomaly, or the anomalous Jacobians
under dilation.  The dilation is nothing but the change in the overall
mass scale:
\begin{equation}
  \phi(x) \rightarrow \phi'(x) = e^{d \lambda}\phi(e^{-\lambda}x) ,
\end{equation}
where $d$ is the canonical dimension of the field: $d=1$ for
Klein--Gordon or vector fields and $d=3/2$ for spinor fields. The
corresponding current is
\begin{equation}
  j_D^\mu = x_\nu \theta^{\mu\nu}
\end{equation}
where $\theta^{\mu\nu}$ is the symmetric (improved) energy momentum
tensor \cite{CCJ}.  The classical Lagrangians with no dimensionful parameters
have invariance under the dilation, while quantum mechanically the
presence of the cutoff destroys the scale invariance, and there is a
trace anomaly,
\begin{equation}
  \partial_\mu j_D^\mu = \theta^\mu_\mu \neq 0.
\end{equation}
The infinitesimal dilation can be written as
\begin{equation}
  \delta \phi(x) = \lambda (d - x^\mu \partial_\mu) \phi(x).
\end{equation}

For scalar fields, the regularized Jacobian for an infinitesimal
dilation is then given by\footnote{Traditionally, the anomalous
  Jacobians under dilation were discussed in terms of Weyl
  transformations \cite{Fujikawa-Weyl}.  We do not use this method
  here to avoid going into supergravity extension of the Weyl
  transformations.}
\begin{equation}
  \ln J = \lambda \mbox{Tr} \left((d - x_\mu
    \partial_{x}^\mu)e^{t(D_\mu)^2}\right)= 
  \mbox{Tr} \left((d - 2 -\frac{1}{2} \{x_\mu, \partial_{x}^\mu\})
    e^{t (D_\mu)^2}\right) 
\end{equation}
where we have used $- x_\mu \partial_{x}^\mu = -1/2[x_\mu,
\partial_{x}^\mu] -1/2\{x_\mu, \partial_{x}^\mu\} = -2 -1/2 \{x_\mu,
\partial_{x}^\mu\}$. Note that, as remarked in the previous appendix,
in the case with constant background electric and magnetic fields, we
found a gauge with $x^\mu A_\mu = 0$, so that in fact the operator
$x^\mu \partial_\mu = x^\mu D_\mu$ appearing in the Jacobian is gauge
covariant.

It is easy to see that the anti-commutator piece does not contribute
to the trace in the Jacobian. Since the eigenstates of $(D_\mu)^2$ (in
the constant $E,B$ background we are considering) are harmonic
oscillator modes, it suffices to note that $\{x_\mu,
\partial_{x}^\mu\} \sim i (a^2_\mu - a^{\dagger 2}_{\mu})$, and so
$\langle n_L,n_R|\{x_\mu, \partial_{x}^\mu\} |n_L,n_R\rangle=0$ for
the $|n_L,n_R\rangle$ harmonic oscillator eigenstates. Therefore,
\begin{equation}
  \ln J = (d-2) \mbox{Tr} e^{t (D_\mu)^2}.
\end{equation}
In other words, the anomalous Jacobian under a dilation for individual
component is exactly the same as under a rescaling, with weight
$(d-2)$.  This can shown to be true for the spinor and vector fields
as well.

The anomalous Jacobian under the dilation can be now easily worked out
for a chiral multiplet in a gauge-field background.  It is given by
\begin{equation}
  \ln J = \lambda \left( - \mbox{Tr}_{\phi} e^{t (D_\mu)^2}
                + \frac{1}{2} \mbox{Tr}_{\psi} e^{-t {\not D}^2}
                \right) ,
\end{equation}
since the auxiliary component $F$ has a canonical dimension $d=2$, and
hence has a vanishing weight $d-2=0$.  Using the formulae given in the
previous appendix,
\begin{equation}
  \ln J = \lambda \frac{-1}{16\pi^{2}} \int d^{4}x
                E B \frac{1-\cosh t(E\pm B)}{\sinh t E \sinh t B}
  = \lambda \frac{1}{16\pi^2} \int d^4 x \frac{1}{2} (E\pm B)^2 + O(1/M^4).
\end{equation}
In supersymmetric notation for a general chiral multiplet, it is
\begin{equation}
  \ln J = \lambda \frac{1}{16} \int d^4 x d^2 \theta
  \frac{t_2(i)}{8\pi^2} WW + O(1/M^4).
\end{equation}
This Jacobian gives the correct one-loop contribution to the
holomorphic $\beta$-function from the chiral multiplet.

In this analysis, the holomorphy between $U(1)_R$ transformation and
dilation is manifest.  The $U(1)_R$ transformation with charge $2/3$
for the chiral superfield $\Phi$ rotates the phases of component
fields with charges $2/3$ for $\phi$, $-1/3$ for $\psi$ and $-4/3$ for
$F$.  The Jacobian is therefore
\begin{eqnarray}
  \ln J &=& i\alpha \left(\frac{2}{3} \mbox{Tr}_{\phi} e^{t (D_\mu)^2}
                - \frac{1}{3} \mbox{Tr}_{\psi} e^{-t {\not D}^2}
                - \frac{4}{3} \mbox{Tr}_{F} e^{t (D_\mu)^2}
                \right) \nonumber \\
        &=& i\alpha \left(-\frac{2}{3} \mbox{Tr}_{\phi} e^{t (D_\mu)^2}
                - \frac{1}{3} \mbox{Tr}_{\psi} e^{-t {\not D}^2}
                \right) ,
\end{eqnarray}
where we used the equality of the traces on $\phi$ and $F$ components.
This is precisely the same as the Jacobian under the dilation except a
factor of $i2/3$ and $\lambda \rightarrow \alpha$.  Note that the form
above is not $t$-independent, but the combination $\ln J + \ln
\bar{J}$ is, and hence the $U(1)_R$ anomaly is exact.  The $F$-terms
in the Jacobians are exact individually for $J$ and $\bar{J}$, as can
be seen by employing the instanton background $E=B$,
\begin{equation}
  \left. E B \frac{1-\cosh t(E \pm B)}{\sinh t E \sinh t B}\right|_{E=B}
  = - \frac{1}{2} (E\pm B)^2 ,
\end{equation}
with no $t$-dependence and given only by the zero modes.  

One can go through the same calculation for a vector multiplet around
each point in the functional space.  In Wess--Zumino gauge, the
contributions come from the all four components of $V_\mu$ after gauge
fixing with weight $-1$, Faddeev--Popov ghosts $c$ and $\bar{c}$ with
weights $-1$ but with the opposite sign, and gauginos $\lambda$ and
$\bar{\lambda}$ with $-1/2$ but with the opposite sign.  Note that
auxiliary fields have vanishing weights and hence do not contribute.
We find
\begin{eqnarray}
\lefteqn{
  \ln J } \nonumber \\
& = & \lambda \left( - \mbox{Tr}_{V} e^{t ((D_\mu)^2\delta_{\mu\nu} 
      -i F_{\rho\sigma}(M^{\rho\sigma})_{\mu\nu})}
    + \mbox{Tr}_{c} e^{t (D_\mu)^2}
    + \mbox{Tr}_{\bar{c}} e^{t (D_\mu)^2}
    + \frac{1}{2} \mbox{Tr}_{\lambda} e^{-t {\not D}^2}
    + \frac{1}{2} \mbox{Tr}_{\bar{\lambda}} e^{-t {\not D}^2}
  \right) \nonumber \\
  & =& \lambda \frac{-1}{16\pi^{2}}\int d^4 x E B 
  \frac{2(\cosh 2tE + \cosh 2tB) - 2 - \cosh t(E+B) - \cosh
      t(E-B)}{\sinh t E \sinh t B} \nonumber \\
    & = & \lambda \frac{-1}{16\pi^2} \int d^4 x 3 (E^2 + B^2) + O(1/M^4).
\end{eqnarray}
Again the instanton background gives a $t$-independent result
\begin{equation}
  E B 
  \frac{2(\cosh 2tE + \cosh 2tB) - 2 - \cosh t(E+B) - \cosh
      t(E-B)}{\sinh t E \sinh t B} = 6 E B ,
\end{equation}
and hence the Jacobian is exact for $F$-terms but not for
higher-dimensional $D$-terms.

It is interesting to compare the above Jacobian with that under
$U(1)_R$ transformation.  They agree only up to higher dimension
$D$-terms.  The Jacobian under the $U(1)_R$ current is simply that
from $\lambda$ with charge $+1$ and $\bar{\lambda}$ with charge $-1$,
and hence
\begin{eqnarray}
  \ln J &=& i\alpha \left( \mbox{Tr}_{\psi} e^{-t {\not D}^2}
                - \mbox{Tr}_{\bar{\psi}} e^{-t {\not D}^2}
                \right) \nonumber \\
        &=& i\alpha \int d^4 x E B \frac{2(\cosh t(E+B) - 2\cosh
                t(E-B))}{\sinh t E \sinh t B} 
        = i\alpha \int d^4 x 4 E B \nonumber \\.
\end{eqnarray}
This is again $i2/3$ times that of the trace anomaly.  One little
surprise here is that the relation between the trace anomaly and
$U(1)_R$ anomaly is not exact, but appears to hold only for finite
pieces.  This is not a true statement because we do not use a
manifestly holomorphic formalism for the vector multiplet.  Recall
that the total $U(1)_R$ anomaly for a chiral multiplet had a
cancellation of higher-dimension operators between $J$ and $\bar{J}$.
Even though the chiral Jacobian $J$ preserves manifest holomorphy
between $U(1)_R$ and trace anomalies, the total Jacobian $J\bar{J}$
does not. The
apparent mismatch between the $U(1)_R$ and trace anomalies in the
explicit forms of the Jacobians is an artifact of the
formalism.  In the $N=4$ regularization of pure SUSY YM we presented
in Sec. 5, the holomorphy between $U(1)_R$ and trace anomalies was manifest.

A manifestly supersymmetric formalism requires three sets of ghost
chiral superfields, $b$, $c$ and $c'$ as reviewed in A.4.  The
contribution from the 
ghost fields is the same as for a chiral multiplet with an overall
multiplicative factor $-3$.  In this formalism the holomorphy between
$U(1)_R$ anomaly and trace anomaly is manifest as well for the
contributions from the ghost chiral superfields.  There is no
contribution from the full vector multiplet to the $U(1)_R$ anomaly.
There is, however, a contribution from the full vector multiplet to
the trace anomaly with the following weights: $C(-2)$, $\psi(-3/2)$,
$\bar{\psi}(-3/2)$, $V_\mu(-1)$, $M(-1)$, $N(-1)$, $\lambda(-1/2)$,
$\bar{\lambda}(-1/2)$ and $D(0)$.  One finds
\begin{eqnarray}
\lefteqn{
  \ln J_V } \nonumber \\
&=& \lambda \frac{-1}{16\pi^2} \int d^4 x E B
  \frac{4 + 2(\cosh 2tE + \cosh 2tB)-2(\cosh t(E-B)+\cosh t(E+B))}
  {\sinh t E \sinh t B} \nonumber \\
&=& \lambda \frac{-1}{16\pi^2} \int d^4 x \frac{(E^2 - B^2)^2}{M^4} +
  O(1/M^8) . 
\end{eqnarray}
Therefore, the leading contribution is a higher dimensional $D$-term
$\int d^4 \theta (WW) (\bar{W}\bar{W})/M^4$ where $M$ is the
ultraviolet cutoff, which can be dropped when one studies the running
gauge coupling constant.  The above combination trivially vanishes
under an instanton background $E=\pm B$.  

The final answer for the Jacobian of the vector multiplet for an
arbitrary gauge group is given in supersymmetric notation by
\begin{equation}
  \ln J = \lambda \frac{1}{16} 
  \int d^4 x d^2 \theta \frac{-3t_i(A)}{8\pi^2} W W + h.c. + O(1/M^4) .
\end{equation}
This Jacobian gives the correct one-loop contribution to the
holomorphic $\beta$-function from the vector multiplet.

\end{document}